%% file: BoutsiaAXAF.tex
\documentclass{aa}
\usepackage{natbib}
\bibpunct{(}{)}{;}{a}{}{,}  
\usepackage[figuresright]{rotating}
\usepackage{graphicx}
\usepackage{txfonts}
\begin{document}

\title{Spectroscopic follow-up of variability-selected active galactic nuclei in the Chandra Deep Field South
\thanks{Based on observations collected at the European Southern Observatory, Chile, 080.B-0187(A)}}


\author{K. Boutsia \inst{1,2,3}, B. Leibundgut \inst{2,3}, D. Trevese \inst{1}, F. Vagnetti \inst{4} }

   \offprints{Konstantina Boutsia, \email{kboutsia@oa-roma.inaf.it}}

   \institute
       {Dipartimento di Fisica, Universit\`a di Roma ``La Sapienza'', P.le A. Moro 2, I-00185 Roma, Italy               
         \and
	 European Southern Observatory, Karl-Schwarzschild-Strasse 2, Garching D-85748, Germany
	  \and
	  Excellence Cluster Universe, Technische Universit\"{a}t M\"{u}nchen, Boltzmannstr. 2, Garching D-85748, Germany
           \and
	   Dipartimento di Fisica, Universit\`a di Roma ``Tor Vergata'', via della Ricerca Scientifica 1, I-00133, Roma, Italy
             }   

\date{}

\abstract
{Supermassive black holes with masses of $10^5 - 10^9 M_{\odot}$ are believed to inhabit most, if not all, nuclear regions of galaxies, and both observational 
evidence and theoretical models suggest a scenario where galaxy and black hole evolution are tightly related. Luminous AGNs are usually selected by their non-stellar 
colours or their X-ray emission. Colour selection cannot be used to select low-luminosity AGNs, since their emission is dominated by the host galaxy. 
Objects with low X-ray to optical ratio escape even the deepest X-ray surveys performed so far. In a previous study we presented a sample of candidates selected through 
optical variability in the Chandra Deep Field South, where repeated optical observations were performed in the framework of the STRESS supernova survey.}
{ The analysis is devoted to breaking down the sample in AGNs, starburst galaxies, and low-ionisation narrow-emission line objects, 
to providing new information about the possible dependence of the emission mechanisms on nuclear luminosity and black-hole mass, and eventually studying the 
evolution in cosmic time of the different populations.}
{We obtained new optical spectroscopy for a sample of variability selected candidates with the ESO NTT telescope. We analysed the new spectra, together 
with those existing in the literature and studied the distribution of the objects in $U-B$ and $B-V$ colours, optical and X-ray luminosity, and variability amplitude. }
{A large fraction (17/27) of the observed candidates are broad-line luminous AGNs, confirming the efficiency of variability in detecting quasars. 
We detect : i) extended objects which would have escaped the colour selection and ii) objects of very low X-ray to optical ratio, in a few cases without 
any X-ray detection at all. Several objects resulted to be narrow-emission line galaxies where variability indicates nuclear activity, 
while no emission lines were detected in others. Some of these galaxies have variability and X-ray to optical ratio close to active galactic nuclei, 
while others have much lower variability and X-ray to optical ratio. This result can be explained by the dilution of the nuclear light due to the host galaxy. }
{Our results demonstrate the effectiveness of supernova search programmes to detect large samples of low-luminosity AGNs. A sizable fraction of the AGN 
in our variability sample had escaped X-ray detection (5/47) and/or colour selection (9/48). Spectroscopic follow-up to fainter flux limits is strongly encouraged.}

\authorrunning{K. Boutsia et al.}
\titlerunning{Variability-selected AGNs}
\maketitle

\section{Introduction}

Large samples are required to perform statistical studies of the population of active galactic nuclei (AGNs) and its evolution.
The general principle for discovering AGNs is to use the characteristics that differentiate them from stars or 
galaxies. The spectral energy distributions (SEDs) of AGNs are remarkable for their broad extent in frequency, which is much wider than for normal galaxies.
The UV/optical emission-line spectra stand out for the strong emission lines and for the high level of ionisation. 
Another distinguishing factor is the compact, point-like structure of the nucleus. 
Historically, AGNs were first discovered as the optical counterpart of radio sources. Radio-loud sources, however, make up only 10\% of the known AGN population \citep{whit00}. 
Currently, the discovery of AGNs by hard ($>$2 keV) X-ray observations is the most straightforward method, since stars and galaxies have weak emission 
at these wavelengths and hard X-rays are less affected by dust obscuration.
Since the selection effects vary in the different 
wavelength bands, the properties of the selected objects may also be different and this can lead to 
discrepancies in
the cosmological evolution of the population as derived from different samples.
This is true, in particular, for the low-luminosity part of the luminosity function (LF) of AGNs.
Most optical samples are constructed by selecting candidates on the basis of their non stellar colour.
Since variability is a property shared by most AGNs, a complementary method for selecting 
AGN candidates consists in detecting all variable objects in the field (notice that, in deep surveys, variable stars represent a small fraction). 
Variability is an optical, yet colour-independent selection technique that has been successfully used in the past \citep[e.g.][]{hawk83,trev89,vero95,trev94,bers98,geha03,sesa07} to select AGN candidates and possibly check the completeness of colour-selected samples. 
This comparison makes sense only for QSOs or bright AGNs, where the nuclear light dominates the SED, while it is known that the colour selection technique is only effective 
above an absolute nuclear magnitude, in the R band, $M_R \sim $ -21.5, since nuclei of lower luminosity are swamped by the light of the host galaxy. This is why the LF of low 
luminosity AGNs (LLAGNS), and its evolution in cosmic time are still poorly known. 
For this reason \citet{bers98} applied for the first time a detection criterion based on variability to objects with extended images, creating a sample of candidates 
in the Selected Area 57, which was subsequently studied in X-rays \citep{trev07} and follow-up optical spectroscopy \citep{trev08b}.
Variability-selected samples were created on the basis of high spatial resolution Hubble Space Telescope Images by \citet{sara03,sara06}, to minimise the dilution 
effect of the (constant) galaxy light.
These studies discovered a number of AGN candidates not detected in X-rays.

We presented another variability selected sample in \cite{trev08a} (hereinafter paper I), created on the basis of a new analysis of the data of the STRESS 
survey \citep{capp05,bott08}, which is a project devoted to the discovery of all supernova (SN) types.
We selected the AXAF field of the STRESS project (after the name of the X-ray satellite, subsequently named Chandra), which is centred at $\alpha$=03:32:23.7, $\delta$=-27:55:52 (J2000). 
We decided to keep the name AXAF adopted by the STRESS team, since the field, which is $33\times34 arcmin^2$ wide, does not simply correspond to the Chandra Deep Field South (CDFS), 
but includes it and is about 3 times larger (see Fig.1 of paper I).
The choice  of the field was motivated by the fact that CDFS is one of the best studied areas in the sky and there is a wealth of information across the electromagnetic spectrum, 
from X-rays to radio. This includes, imaging in 17 bands performed by the COMBO-17 survey \citep{wolf03}, 
from which low-resolution SEDs have been obtained for the classification of the sources in stars, galaxies and QSOs. 
There is the ESO Imaging Survey \citep[EIS,][]{arno01}, which provides photometry in 5 bands and size 
information for the sources. Two X-ray surveys have been conducted in this area, the Chandra Deep Field South 
\citep[CDFS,][]{giac02,alex03} with exposure times of 1Ms (recently re-observed for another Ms \citep{luo08}), and the 
Extended-CDFS survey \citep[ECDFS,][]{lehm05}, of 250ks. This area has been targeted by several spectroscopic campaigns
\citep{szok04,le-f04,graz06,ravi07,pope08,trei08}. 
Recently, a catalogue of radio sources in the same area has been published by \cite{kell08}. 
A variability study has also been performed by \citet{kles07} in the GOODS South field which is contained in the AXAF field analysed in the present study. 
A comparison of our and their results has been presented in paper I. This makes this area ideal to study 
completeness, selection effects and biases of the different methods to identify AGNs. 

Despite the several surveys which were conducted in this area, when we started this analysis, only 25\% of our variable sources (30/132) had 
spectroscopic verification from previous studies. For this reason we have performed spectroscopic follow-up for the bright part of our sample, in order to confirm the 
AGN nature of our candidates and obtain redshifts. After this study, 80\% (36/45) of our candidate sample to a magnitude limit 
of $\sim$21.3 in the V band has a spectroscopic redshift. From our total sample, that reaches magnitude V=24,
only $\sim$55\% (72/132) has optical spectroscopy. Out of these 72 objects, 12 can be considered as low-luminosity sources ($L_{R} < 10^{43} erg s^{-1}$). 
We believe that this number can increase significantly and therefore we will pursue the completion of the spectroscopic follow-up for 
the entire variability selected sample. Here we present the results obtained for a subsample of sources drawn mainly from the bright part of our original 
variability selected candidate list. For these sources, we have available spectroscopic information either by our campaign or from the literature that 
allow us to explore their properties.

The paper is structured as follows. Section 2 describes the object selection, the spectroscopic observations and data reduction procedure, and the 
derived spectra. In section 3 we present the properties of all the variable candidates with optical spectra in our sample. In Section 4 we summarise 
our results. Comments on individual objects are presented in the Appendix.

Throughout the paper we adopt the concordance cosmology: H$_{o}$=75 km s$^{-1}$ Mpc$^{-1}$, $\Omega_{m}$=0.3, $\Omega_{\Lambda}$=0.7.

\section{Optical spectroscopy}

\subsection{Selection of candidates and observations}

The AXAF field was observed 8 times in about 2 years with the wide field imager at the ESO/MPI 2.2m telescope. In paper I, light curves were derived for 
each object in the field. Variability amplitude was measured by computing the r.m.s. variations $\sigma$  and variable objects were selected according to the condition:

\begin{equation}
\sigma^*\equiv \frac {\sigma- s(V)}{\Sigma_{\sigma}(V)} \ge 3
\end{equation}

where $s(V)$ and $\Sigma_{\sigma}(V)$ are the ensemble average and standard deviation of $\sigma$, as a function of the magnitude V.
Details about the selection criteria, the complete catalogue of 
the variability selected candidates in the AXAF field, and a comparison of our sample with the already existing ones in the same field 
have been presented in \citet{trev08a}. 
In the present follow-up campaign we simply started from the catalogue of variable objects selected in paper I, we excluded all 
those objects whose redshift was already known from the literature, then we sorted the remaining objects according to increasing $V$ 
magnitude and prepared a list to observe as many as possible, starting with the brightest ones.

The observing run on La Silla took place from the 1st to the 5th of November 2007. We used EMMI/NTT in 
the red imaging and low dispersion spectroscopy (RILD) mode \citep{dekk86} to perform low resolution long-slit spectroscopy. 
We used grism 2 which has a dispersion of 1.74$\AA$/pixel and covers the wavelength range between 380 and 920nm and a 1.0$\arcsec$ slit with 8$\arcmin$ length. 
We did not use an order-sorting filter and there might be second order contamination 
beyond 800nm. The CCD detector is a 2-chip mosaic. The pixel scale is 0.166$\arcsec$/pixel in the 1$\times$1 
binning mode. The field size is 9.9$\arcmin$$\times$9.1$\arcmin$ and the gap between the 2 chips 
is 47pixels wide, which corresponds to 7.82$\arcsec$.

Using a 2$\times$2 binning and the 1.0$\arcsec$ slit, we obtained a resolution $\frac{\lambda}{\Delta\lambda}$=570 which allows us 
to resolve emission lines broader than $\sim$500km/s. This is enough to detect narrow emission lines in AGN. Since our targets are 
faint, in order to observe as many candidates as possible, in each pointing we placed two sources onto the slit. 
Therefore the slit was not positioned at the parallactic angle. However, at the time of exposure, most of our candidates had an airmass 
in the range of 1.0-1.1 and only the pairs towards the end of the night would exceed an airmass of 1.2. Thus, the 
effect of the light losses is negligible. Since our main interest is to measure redshifts and possibly determine ratios 
between neighbouring lines, we accept this limitation. The exposure times ranged from 900s 
for sources with magnitude V$<$20.3 to 1800s for sources with 20.3$<$V$<$21.3. At the beginning of the night we acquired arc 
spectra for wavelength calibration and during the night we observed standard stars for flux calibration.

\begin{figure}
\hspace{-1.0cm}
\includegraphics[width=18cm]{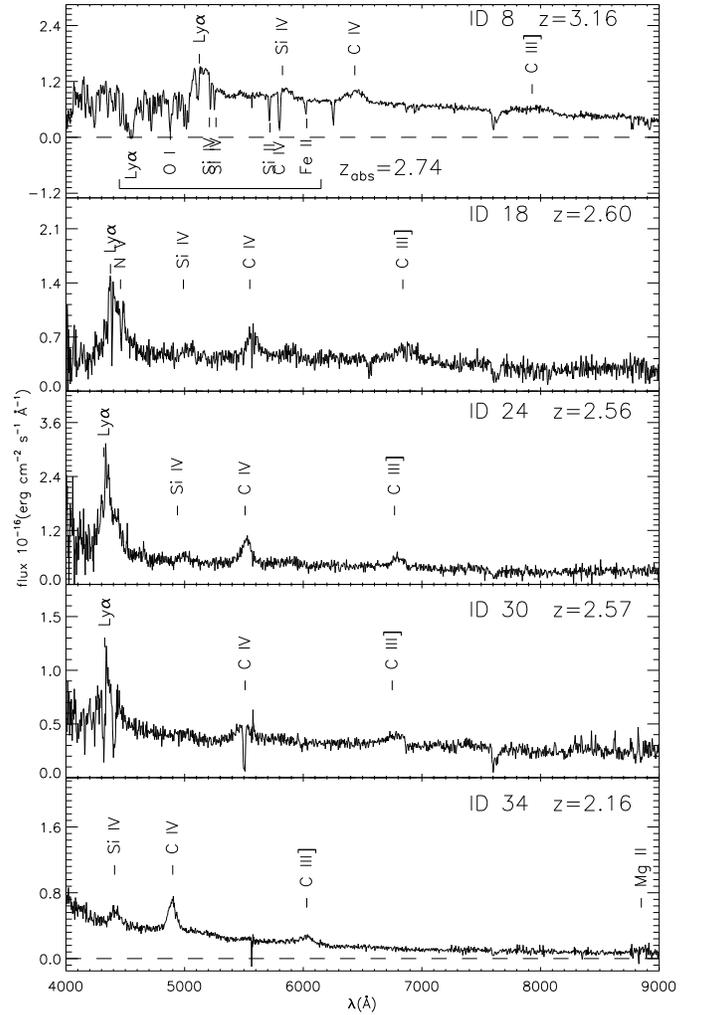}
\vspace{0.05cm}
\caption{\small Spectra of BLAGN that have the Ly$\alpha$ line within the wavelength range ($\emph{z}$ $>$ 2).}
\label{LyA}
\end{figure}

The data reduction was carried out with standard IRAF procedures. Our exposures were long, therefore we used 
the L.A. Cosmic software \citep[Laplacian Cosmic Ray Identification,][]{van-01} to detect and remove cosmic rays from the images. 
The sky subtraction was done by manually selecting the sky region. After calibrating the wavelength scale, we 
performed a flux calibration using the spectra of standard stars. We obtained spectra for 27 sources. The redshift 
measurement is accurate to 0.01 ($\Delta\emph{z} \leq 0.01$). The flux calibrated spectra are shown in Figs. \ref{LyA}-\ref{SF} and \ref{other}.

\input{1092T1.tex}

The detailed list of the sources observed during this run is presented in Table \ref{Tab1}, where 
the following information can be found:
{\it Column 1}: object identification No. (from paper I);
{\it Columns 2 and 3}: right ascension $\alpha$ and declination $\delta$ (J2000);
{\it Column 4}: V magnitude (from paper I);
{\it Column 5}: the maximum magnitude change $V_{min}-V_{max}$;
{\it Column 6}: exposure time;
{\it Column 7}: average airmass at the time of observation;
{\it Column 8}: the spectroscopic redshift we have measured;
{\it Column 9}: the spectral classification of the source.

The morphology of the images as obtained with the ESO/MPI 2.2m telescope, in the V band, are shown in Fig. \ref{thumbs}.

\subsection{Spectra and classification}
There are two main categories of objects. The ones with broad emission lines were classified as broad-line AGNs (BLAGNs). We also detected 
7 sources that have low redshifts and only narrow emission lines, including H$\alpha$. We classify these 
sources as Narrow Emission Line Galaxies (NELG). In these objects, although we always detect the H$\alpha$ emission line, we do not 
always have the H$\beta$ and [OIII]$\lambda$5007$\AA$. This means that we can only derive upper limits for the line 
ratios and thus diagnostic diagrams cannot help us to robustly distinguish the nature of these objects. 

\subsubsection{Objects with $\emph{z}$ $>$ 2}
There are 5 sources with broad emission lines and the redshift determination is based mainly on the 
detection of Ly$\alpha$$\lambda$1216$\AA$, CIV$\lambda$1549$\AA$ and CIII]$\lambda$1909$\AA$. For all 
these objects we have secure redshifts. They are classified as broad-line AGN (BLAGN) and the ones included 
in the ECDFS field have also detected emission in the X-ray band. The spectra are displayed in Fig. 1.

\begin{figure}
\hspace{-0.5cm}
\includegraphics[width=13cm]{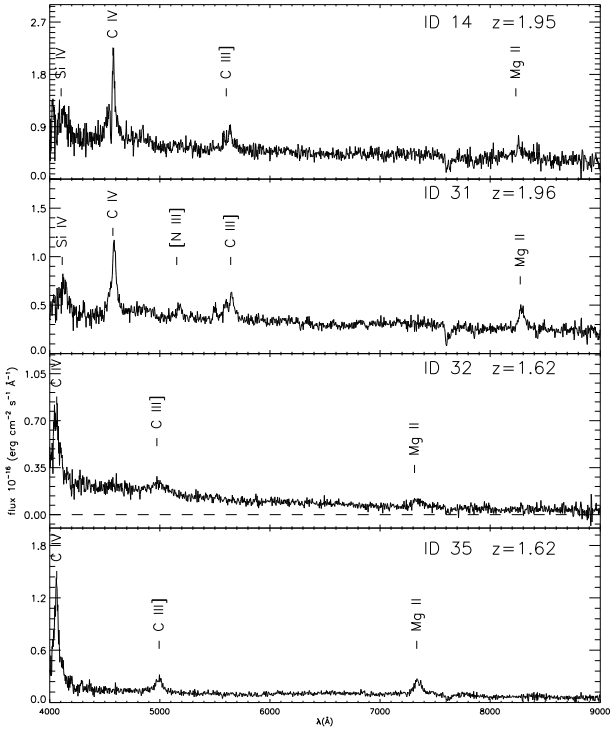}
\vspace{0.05cm}
\caption{\small Spectra of BLAGN in the 1.5 $<$ $\emph{z}$ $<$ 2.0 redshift range.}
\label{CIV}
\end{figure}

\subsubsection{Objects with 1.5 $<$ $\emph{z}$ $<$ 2.0}
For these sources, the redshift determination is based on the detection of CIV$\lambda$1549$\AA$, CIII]$\lambda$1909$\AA$ 
and MgII$\lambda$2798$\AA$ emission lines (see Fig. \ref{CIV}). They all show X-ray emission, 
except for the source ID 31, which lies outside the fields covered by the X-ray surveys. They are all classified as BLAGN.

\begin{figure}
\hspace{-0.5cm}
\includegraphics[width=13cm]{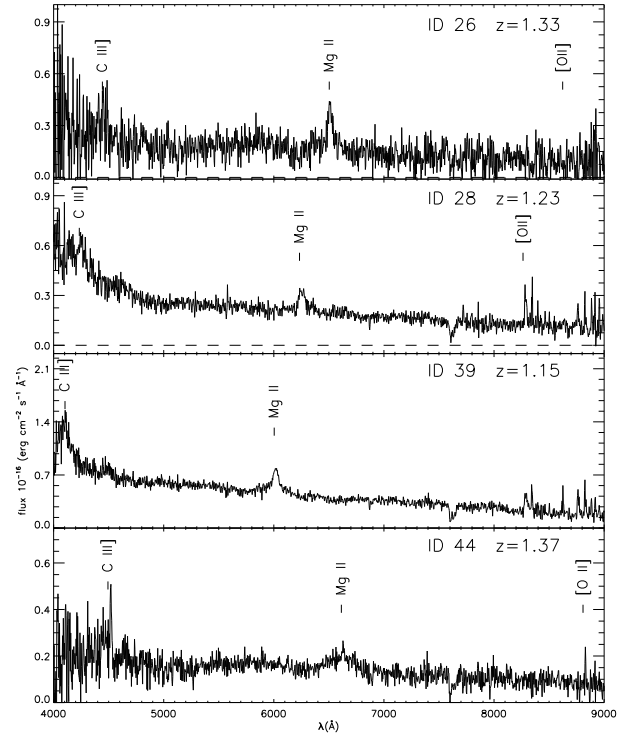}
\vspace{0.05cm}
\caption{\small Spectra of BLAGN in the 1.0 $<$ $\emph{z}$ $<$ 1.5 redshift range.}
\label{CIII}
\end{figure}

\subsubsection{Objects with 1.0 $<$ $\emph{z}$ $<$ 1.5}
For these four sources the redshift is based on the detection of CIII]$\lambda$1909$\AA$ and MgII$\lambda$2798$\AA$ (see Fig. 3). 
In some cases the [OII]$\lambda$3727$\AA$ line is within our spectral range, but since it is at the very edge, it is not always 
prominently detected. Nevertheless, with 2 broad lines in the spectral range, we consider our redshift determination secure and we classify these 
objects as BLAGN. Most of these objects have been classified as galaxies by their SED in COMBO-17 and they are good examples 
of the kind of sources that the variability selection can bring to light. In particular the source ID 26 is of great interest, since 
it has not been detected in the X-rays, even though it is within the area of ECDFS and it was classified as galaxy by COMBO-17. Based on ECDFS 
intesity maps the upper limit for the flux in the hard band is ${f}_{X}(2-8keV) < 1.77~10^{-15}~erg~cm^{-2}~sec^{-1}$ (for more details see the Appendix).

\begin{figure}
\hspace{-0.5cm}
\includegraphics[width=13cm]{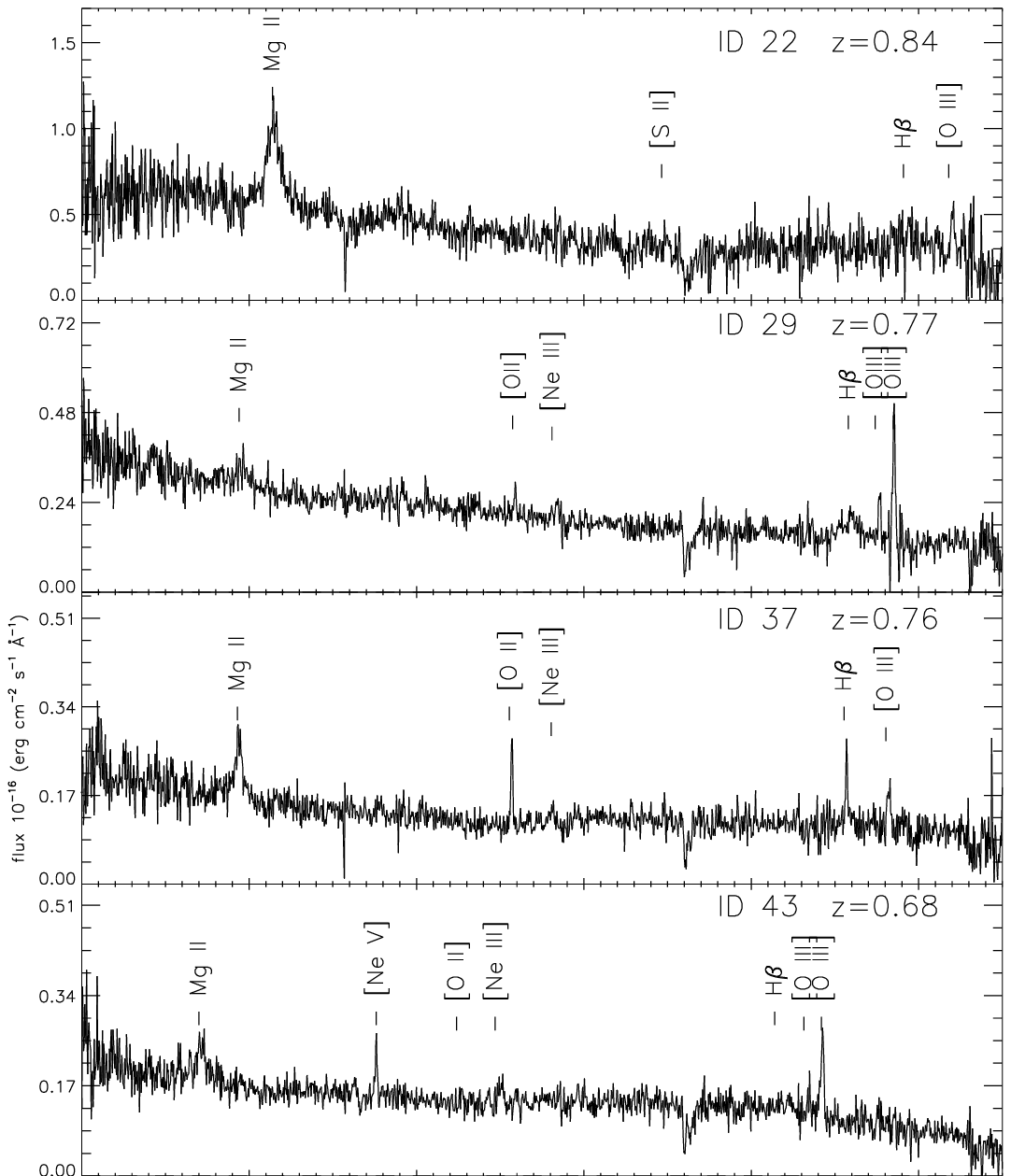}
\vspace{0.05cm}
\caption{\small Spectra of BLAGN in the 0.6 $<$ $\emph{z}$ $<$ 1.0 redshift range.}
\label{MgII}
\end{figure}

\subsubsection{Objects with 0.6 $<$ $\emph{z}$ $<$ 1.0}
These four sources display only one broad emission line. We interpret this to be MgII$\lambda$2798$\AA$. Other emission lines are also 
detected in this spectral range, like [OII]$\lambda$3727$\AA$, H$\beta$ and [OIII]$\lambda$5007$\AA$ (see Fig. 4). They are all 
classified as BLAGN mainly because of the presence of MgII$\lambda$2798$\AA$. Three of these sources (ID 22, ID 37, ID 43) show X-ray emission, 
while the other one (ID 29) is outside the area covered by Chandra (for more details see the Appendix).

\subsubsection{Objects with narrow emission lines (NELG)}
These candidates exhibit narrow emission lines, like the H$\alpha$$\lambda$6563$\AA$, [NII]$\lambda$6584$\AA$ doublet and 
the [SII]$\lambda$$\lambda$6717,6731$\AA$$\AA$ doublet, with no detectable broad component. Most of the times the 
forbidden [OII]$\lambda$3727$\AA$ line is present within the spectral range (see Fig.5). The redshift determination is secure, 
but since H$\beta$ and the [OIII]$\lambda$$\lambda$4959,5007$\AA$$\AA$ doublet are not always visible we can not distinguish between star-forming 
galaxies or faint AGN and unambiguously determine their nature. However, the fact that, when detected, the [OII]$\lambda$3727$\AA$ line 
is very prominent, while the other lines are marginally visible, would tend to suggest that they are star-forming galaxies. 
This is also supported by the fact that they don't exhibit the power-law continuum, typical of AGN. In this case another issue 
emerges, because starburst galaxies are not supposed to be significantly variable, yet almost all these candidates 
have $\sigma$*$>$3.5. The origin of their variability is an interesting problem and such objects are worth further 
investigation. Since no secure classification is possible, we generally define these objects as Narrow Emission Line 
Galaxies (NELG). For the objects of this class that are within the fields covered by the X-ray surveys, no X-ray emission 
is detected and we calculate upper limits based on the ECDFS intensity maps. 

In an effort to better constrain the nature of these sources we calculate the line ratios and we plot them on the diagnostic 
diagram that can be seen in Fig.6. The classification of the sources follows the work of \cite{kewl06}, from where we have taken 
the equations that distinguish the different areas. As it is seen from the spectra in Fig.5, we don't always have all lines 
necessary, but nevertheless we calculate the ratios, considering the flux value of the continuum as a lower limit of the emission 
line flux. We see that two sources (ID 11, ID 12) are in the ``composite objects'' area and this is a strong indication for the presence 
of a faint AGN. Apart from the object ID 15, the rest of the sources have only lower limits for their ratio and we cannot robustly 
claim the presence or absence of an active nucleus.  

\begin{figure}
\hspace{-1.2cm}
\includegraphics[width=20cm]{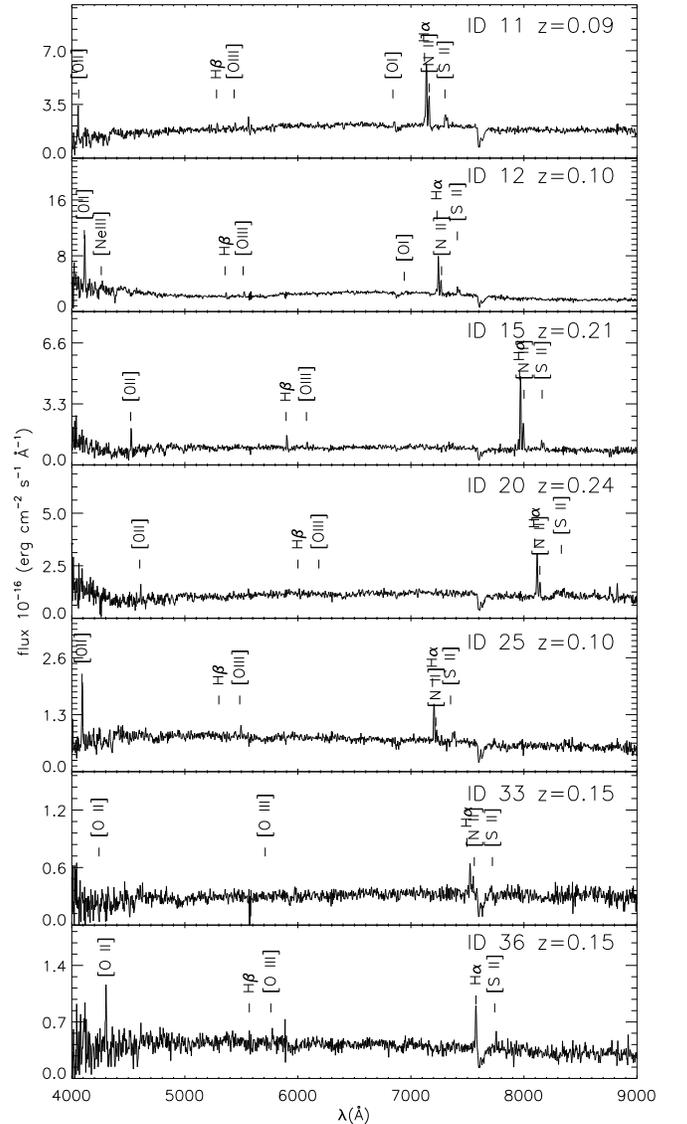}
\vspace{0.05cm}
\caption{\small Spectra of the sources classified as Narrow Emission Line Galaxies (NELG).}
\label{SF}
\end{figure}

\begin{figure}
\hspace{-2cm}
\includegraphics[width=12cm]{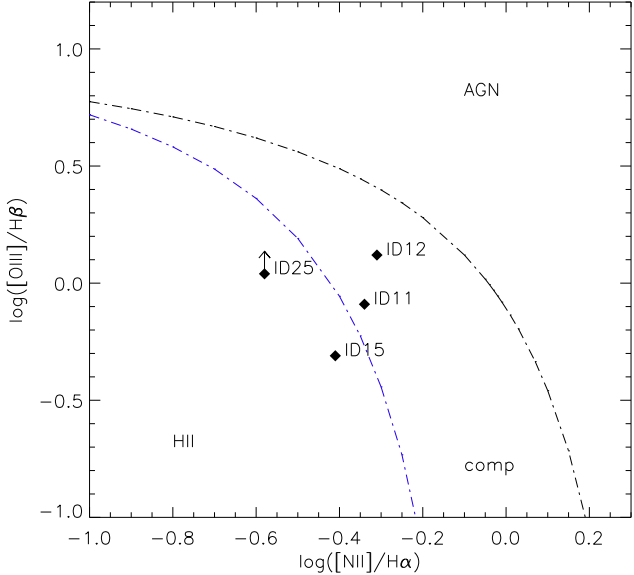}
\caption{\small Diagnostic diagram for the NELG sources. The curves separate the areas where the different objects are found according 
to the classification scheme presented by \cite{kewl06}.}
\label{diagn-dia}
\end{figure}

\begin{figure}
\hspace{-0.5cm}
\includegraphics[width=12cm]{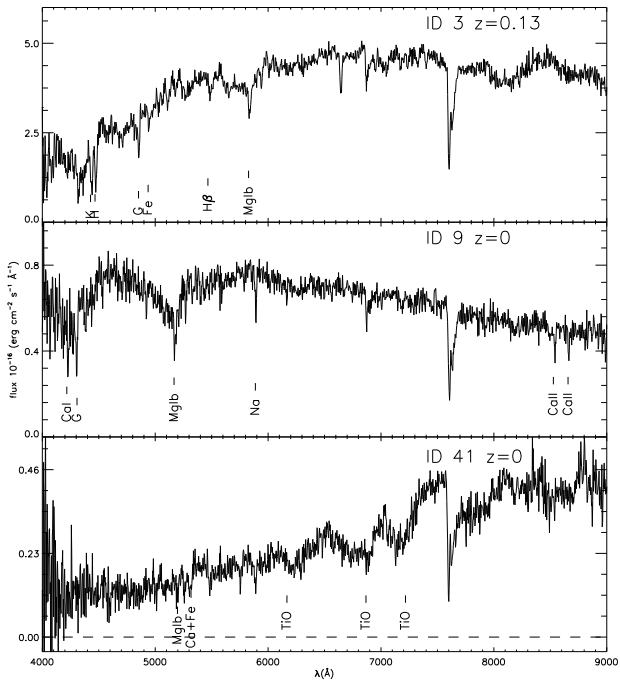}
\vspace{0.05cm}
\caption{\small Spectra of the normal galaxy (upper panel) and the two stars (middle and lower panel) that were observed during our follow-up campaign.}
\label{other}
\end{figure}

\subsubsection{Other}
Three objects do not show emission lines and the redshifts have been determined based on the absorption features. The two stars (ID 9, ID 41) 
have been classified as such also by \cite{groe02} and appear in the 5-passband stellar catalogue. The source ID 3 turned out to be an absorption 
line galaxy, although normal galaxies are not expected to be variable.  

\section{Properties of sources with optical spectra}

The new spectra presented in this work, together with the spectroscopic information available from the literature, allow us to characterise 
variability-selected sources. The redshifts from the literature come mainly from \citet{szok04} and also from \cite{trei08,ravi07,graz06,le-f04}. 
All relevant information is assembled in Table \ref{Tab2}, where we report:
{\it Column 1}: object identification No. (from paper I);
{\it Columns 2 and 3}: right ascension $\alpha$ and declination $\delta$ (J2000);
{\it Column 4}: V magnitude (from paper I);
{\it Column 5}: standard deviation $\sigma$ of the light curve, taken from paper I;
{\it Column 6}: normalised standard deviation $\sigma^*$ (Eq.(1));
{\it Column 7}: EIS name;
{\it Column 8 and 9}: $V$ and R magnitudes, respectively, from the EIS catalogue (AB system);
{\it Column 10 and 11}: $U-B$ and $B-V$ colours, respectively, from the EIS catalogue (AB system);
{\it Column 12}: stellarity index in the V band from the EIS catalogue;
{\it Column 13}: COMBO-17 name;
{\it Column 14}: COMBO-17 class;
{\it Column 15}: spectroscopic redshift; 
{\it Column 16}: spectroscopic classification;
{\it Column 17}: X-ray identification from \cite{lehm05} except otherwise noted;
{\it Column 18}: X-ray flux (2-8keV) in erg cm $^{-2}$ s$^{-1}$;
{\it Column 19}: logarithm of the X-ray luminosity in erg s$^{-1}$;
{\it Column 20}: logarithm of R band luminosity in erg s$^{-1}$;
{\it Column 21}: logarithm of the hard (2-8keV) X-ray flux to optical (R band) flux ratio;
{\it Column 22}: notes.

For the purpose of this study it is convenient to separate objects in three groups, 
according to their spectral features, and establish the symbols that will be adopted in all of the following figures. 
We consider: i) broad-line AGNs (BLAGNs), the sources with at least one broad emission feature in the observed spectral range and 
they will be represented by filled circles; ii) narrow emission line galaxies, the sources with low excitation emission lines and 
without apparent broad features in the observed spectral range. They may include starburst galaxies and LINERs and will be represented by diamonds; 
iii) galaxies, the objects where only absorption features are detected and are represented by squares 
(the above symbols, in the electronic version, have different colours: black, red and blue, respectively). 
Crosses indicate that the object possesses an X-ray flux measurement in the ECDFS \citep{lehm05} or CDFS \citep{giac02} catalogues. We have a total of 72 candidates 
with spectroscopic redshift and 27 of these come from our spectroscopic follow-up. We have detected 17 broad-line AGNs (BLAGNs) and 7 Narrow Emission 
Line Galaxies (NELG). At least 2 of the NELG (ID 11, ID 12), according to the diagnostic diagram (see Fig. 6), could be low-luminosity AGN. If we consider 
as genuine AGN only the ones that exhibit broad lines, we have a lower limit of 71\% (51/72) as far as reliability is concerned. In Table \ref{Tab2} we present 
only the extragalactic sources, thus omitting the 2 stars detected in this follow-up. We also omit the 2 SNe reported in paper I, detected in the AXAF field by the STRESS survey. 

\begin{figure}
\hspace{-2cm}
\includegraphics[width=12cm]{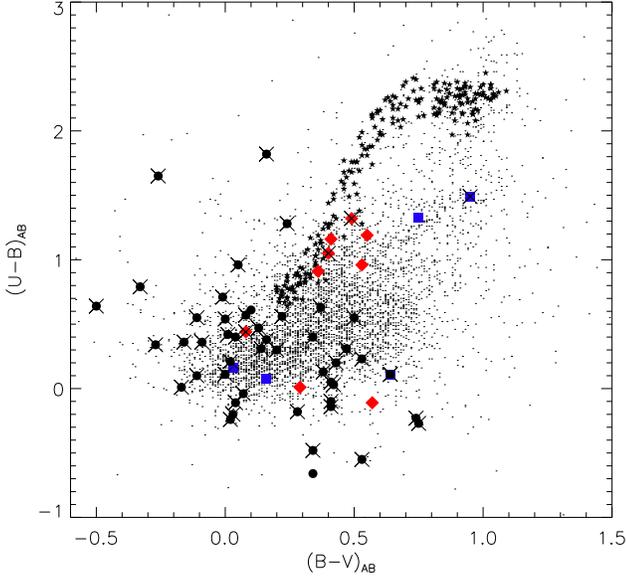}
\caption{\small Colour-colour diagram for the candidates with spectroscopic redshift. Large black dots: BLAGN; red diamonds:NELG; blue squares: galaxies. 
The crosses correspond to sources with X-ray detection. Small star symbols indicate the stellar locus and small dots are non-variable objects.}
\label{CC}
\end{figure}

In Fig. \ref{CC} we present the (U-B)$_{AB}$ vs. (B-V)$_{AB}$ diagram for the candidates with spectroscopic redshift and colour from the EIS catalogue, 
where the stars in the field (small star symbols) are shown to identify the sequence of stellar colours, usually adopted to identify
the ``non-stellar'' pointlike objects as QSO candidates. The small dots represent non variable objects, most of which are galaxies lying  
outside the stellar locus. This is the reason why the colour selection technique is restricted to point-like objects. The majority of variable 
candidates, included in the field of the X-ray surveys, are X-ray emitting objects (46/61), and most of them are broad-line AGN (42/46). 
This is expected in the framework of the standard unification theory, since in type 2 objects the variable nuclear radiation is obscured 
by the absorbing torus. Notice, however, that in some cases objects selected through variability show only narrow lines consistent with 
type 2 AGNs (e.g. the objects NSER 4326 and NSER 16338 discussed in \cite{trev08b}), possibly connected with a variability of the spectrum, 
observed in a particular phase when the broad-line component is absent or reduced. 

Type 1 objects (filled circles) have the characteristic distribution of QSOs. Most of them show the typical UV excess, except for the ones with higher redshift that are ``redder" 
in $(U-B)$ and lie on the left of the stellar locus (eg. ID 30, ID 115, ID 40 ID 120 with z = 2.57, 2.726, 2.81, 2.796 respectively).
Most of these objects would have been selected on the basis of their colour, since most of them : i) are pointlike; ii) have non-stellar colours. However, 
5 broad-line objects (ID 26, ID 74, ID 87, ID 94, ID 125) are not detected in hard X-rays (2-8keV), despite the depth of the Chandra survey. These objects have upper limits 
between 1.3-1.8$\times$10$^{-15}$ erg cm$^{-2}$ s$^{-1}$ and represent 11\% of the variable BLAGN in the area with X-ray information. This is of particular interest 
because, in this case, obscuration cannot be evoked to explain the lack of X-ray emission.  

The variable candidates which turned out to be narrow line objects, can be divided in two groups: with $(U-B)_{AB}<0.5$ (3 sources) or $(U-B)_{AB}>0.8$ (6 sources). 
Among the former three, one (ID 78) has all the characteristics of a normal QSO (colour, variability, X-ray emission, and point-like image), but its 
spectrum has been classified as ``narrow line" by \citet{szok04} (see, however, the discussion below). The other two objects (ID 121, ID 130) are not detected in 
X-rays, in both soft and hard bands, and have extended images. Among the 6 objects classified as NELGs and having $(U-B)_{AB}>0.8 $, three 
(ID 4, ID 5, ID 11) are detected in X-rays. They are mentioned also as extra-nuclear X-ray sources by \cite{lehm06}, that consider as such, the X-ray 
sources observed within the optical extent of the bright galaxy but with an offset from the optical nucleus. For this reason we explore the possibility 
that the extra-nuclear source can affect our variability measurement. In the case of our candidates the distances from the centre of the optical 
image are $4.7 \pm0.9$, $0.9\pm0.6$ and $3.2\pm1.4$ arcsec, respectively, while the aperture used to measure their optical variability is 0.9 arcsec. 
For objects ID 4 and ID 11 we can confidently assume that variability is not associated with the extra-nuclear source, which, on the other hand, 
is not expected to produce appreciable optical variability. Thus, in these two cases an active nuclear component and the extra-nuclear X-ray source seem 
to cohabit the same host galaxy. The case of ID 5 is more complicated since: i) the extra-nuclear source is closer to the optical nucleus;
ii) variability was not detected in a series of 5 ACS Hubble Space Telescope images, obtained during 6 months, and analysed by \citet{kles07}, despite the 
much smaller aperture (0.075 arcsec) adopted, which reduces the dilution caused by the host galaxy. 
However, as discussed in paper I, the presence of variability in our data and its absence in \citet{kles07} analysis can be simply due to the different epoch 
STRESS and HST observations, and, more important, to the fact that our data refer to a 2 yr time span, compared to 0.5 yr of ACS images.
The other 3 narrow-line objects with $(U-B)_{AB}>0.8$ have only upper limits in the hard X-rays band.
Finally, of the 5 objects without detected emission lines, 3 have $(U-B)_{AB}<0.2$, consistent with late spirals at low redshift
and two have $(U-B)_{AB}>1.3$; one of them (ID 88) is detected in X-rays. 

\begin{figure}
\hspace{-2cm}
\includegraphics[width=12cm]{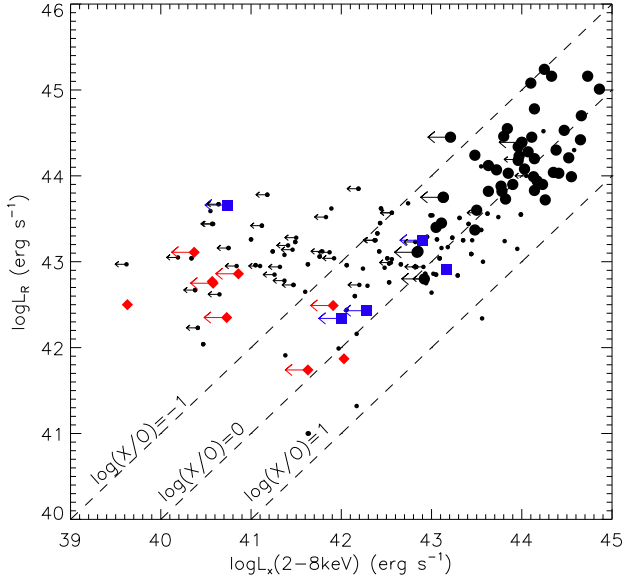}
\caption{\small Luminosity in the R band (logL$_{R}$) vs. luminosity in the hard X-ray band (2-8)keV (logL$_{X}$) for the candidates with spectroscopic redshift. Large black dots: 
BLAGN; red diamonds:NELG; blue squares: galaxies; Small dots: non-variable objects. Arrows indicate 3-$\sigma$ upper limits.}
\label{LRLX}
\end{figure}

Fig. \ref{LRLX} shows the distribution of the optical $L_R$ luminosity versus the X-ray luminosity
$L_X(2-8keV)$ for all the objects in the AXAF field which possess a spectroscopic redshift and are either variable or X-ray detected. Large symbols represent variable 
objects, while small dots represent X-ray detected objects below the variability threshold. We will discuss separately the following three classes of objects: a) broad-line AGNs (BLAGNs); 
b) narrow emission line galaxies (NELGs); c) objects without emission lines (Gal). The straight lines represent constant values of the X-ray (2-8keV) to optical (R band) flux ratio (X/O). 
The luminosities reported are calculated without k-correction. 

Most of the variable sources are broad-line AGNs, occupying a typical region, i.e. the stripe $-1 \la \log(X/O) \la 0.6$ for $\log [L_X(2-8keV)] \ga 42.5 $. 
One exception is the source ID 26, that has been detected through variability, it is in the CDFS field but has 
not been detected in the X-rays and it has no radio emission. After the spectroscopic follow-up, its BLAGN nature was confirmed. 
It has $\log(X/O)\la -1.25$ as upper limit, which is rather low for an AGN, but consistent with similar findings in our previous work in the field of the 
Selected Area 57 \citep{trev08b}. We notice that a fair number of our broad-line AGNs 
were classified as galaxies by the COMBO-17 survey (9/43) (considering as galaxies all objects classified G or G/U in Table \ref{Tab2}).

Among the objects classified as NELGs, we notice that the 3 objects with $(U-B)_{AB}<0.5$ in Fig. \ref{CC}, are the same objects that 
occupy the ``AGN stripe" in Fig. \ref{LRLX}: two of them (ID 121, ID 130) only as upper limit 
on $L_X(2-8keV)$. These two objects have a diffuse appearance ($S_{EIS}= $0.07 and 0.10 respectively),
the third one (ID 78) has a measured X-ray flux, it is pointlike ($S_{EIS}= $0.95) and has been classified as narrow-line by \citet{szok04}, 
but an inspection of the spectrum shows that it is consistent with the presence of a broad MgII($\lambda2798$), at the edge of the blue border 
in the observed spectral range. Thus, this object could correspond to a relatively faint, otherwise normal, BLAGN whose broad line component has not been recognised.
All of the 6 NELGs with $(U-B)_{AB}>0.8$ are also segregated, in the $\log L_R- \log L_X$ plane, but in a region of low $L_X$ and low X/O. 
We stress that, in the case of the sources that have been described also as extra-nuclear, the measures of the X-ray luminosity, or the relevant upper limits, 
refer to the nucleus, and not to the extra-nuclear source.

Concerning the objects without emission lines, 4 of them are consistent with the AGN stripe, while the fifth (ID 3) has a very low $\log(X/O) \approx -3$. Among the former 4, 
one (ID 88) has a measured X-ray flux corresponding to $L_X(2-8keV)\simeq 1.5 \cdot 10^{43}$ erg s$^{-1}$ and can be classified as X-ray Bright Optically Normal Galaxy (XBONG) 
described by \citet{fior00, coma02a,coma02b}. We notice that it has a compact image ($S_{EIS}=0.81$), its $(B-V)_{AB}$ colour ($(B-V)_{AB}= 0.64$) is consistent with an early type galaxy, 
while its $(U-B)_{AB}=0.11$ suggests the presence of an UV emitting nucleus. Its variability also indicates the presence of an AGN. A possible explanation is that we are 
dealing with a normal AGN, not recognised as such because it lies in a redshift interval such that both MgII and H${\alpha}$ lines fall outside the spectral range covered 
by the observations of \citet{le-f04} (i.e. $\lambda \approx (5500-9000)$ \AA), as suggested in similar cases also by \citet{horn05}. As an example, consider our spectrum of ID 22 
in Fig. \ref{MgII}, whose AGN nature would not have been recognised if the spectrum were cut at e.g. $\lambda=5500$\AA.

\begin{figure}
\hspace{-2cm}
\includegraphics[width=12cm]{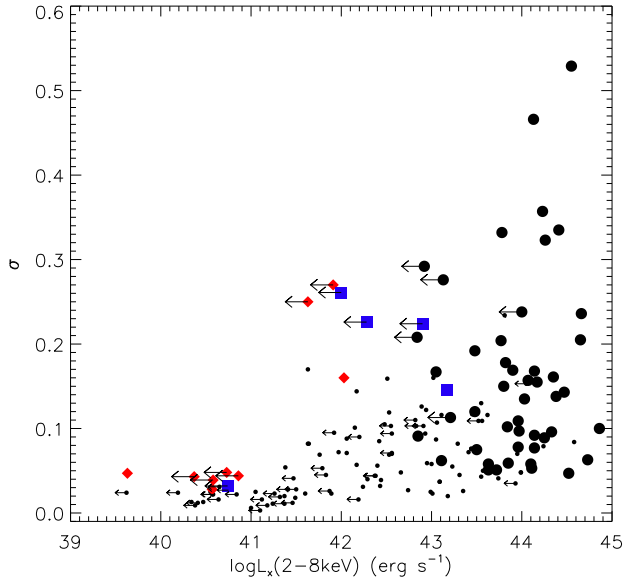}
\caption{\small Variability measurement $\sigma$ vs. logL$_{X}$ in the hard band (2-8)keV. Symbols are as in Fig. 9.}
\label{Svar}
\end{figure}

Finally, in Fig. \ref{Svar} we plot the variability measurement $\sigma$ vs. the logL$_{X}(2-8keV)$ for all objects with redshift. 
Objects below the threshold $\sigma^*= 3$, indicated as small dots in Fig. \ref{LRLX}, have generally a lower variability $\sigma$, as expected.
Notice that some small dots have non-negligible $\sigma$ values, but consistent with photometric noise in the V band at 3$\Sigma_{\sigma}$ (see Eq.(1)). 
It appears from this figure that variability decreases sharply below $L_{X}(2-8keV) \sim 10^{41.5}$ erg s$^{-1}$.
The 6 NELGs with $(U-B)_{AB}>0.8$ and one galaxy without emission lines (ID 33), which show low X/O and low $L_X$ in Fig. \ref{LRLX}, are segregated 
in the $\sigma-\log L_X$ plane, to the bottom left of the diagram. The other 3 NELGs and four galaxies, with higher X/O, have a $\sigma$ comparable with that of 
normal AGNs. Both groups of objects, with stronger or weaker AGN characters respectively, appear segregated consistently both in $L_R-L_X$ and $\sigma-L_X$ planes, 
supporting the claim that their behaviour is due to real physical homogeneity.

One possible explanation for the behaviour of the two groups can be the dilution of the nuclear light by the host galaxy emission.
In fact, if we ascribe entirely to the nuclear component both variability and X-ray emission, we can consider the effect of the host galaxy on the observed variability 
and the X/O ratio as follows. Denoting with $L=L_N+L_g$ the total luminosity, in some optical band, resulting from the nuclear luminosity $L_N$ and the luminosity $L_g$ of 
the host galaxy, we have $\sigma= \sigma_{int} / (1+L_g/L_N)$, where $\sigma=\Delta L_N/ L$ and $\sigma_{int}=\Delta L_N/L_N$ indicate the observed and the intrinsic 
relative flux variations respectively. Similarly the observed X-ray to Optical flux ratio $(X/O)=L_X/(L_N+L_g)$ is diluted with respect to the 
intrinsic $(X/O)_{int}=L_X/L_N$, so that we can write:
\begin{equation}
\frac{\sigma}{\sigma_{int}} \sim \frac{(X/O)}{(X/O)_{int}} \sim \eta \equiv \frac{1}{1+{L_g}/{L_N}}
\quad .\end{equation}

A detailed model is beyond the aim of the present study and would require to take into account not only the k-correction, the effect of the aperture magnitude and the apparent 
variability due to photometric noise, but also the change of $(X/O)_{int}$ and $\sigma_{int}$ with luminosity \citep[see discussion in][]{trev07}.
Here we simply note a qualitative agreement indicating that dilution by the host galaxy can account for the progressive decrease of the average X/O 
below $L_X(2-8keV) \approx 10^{41.5}$ erg s$^{-1}$ and, at the same time, for the relatively sharp decrease in the observed variability at low X-ray luminosity.

For instance, with $\log L_g \approx 42$, the dilution function $\eta$ would fall from $\sim$0.9 to $\sim$0.09 when $\log L_N$ goes from 43.0 to 41.0, 
which would correspond to $\log L_X$ decreasing from 42.5 to 40.5, if we extrapolate to low X-ray luminosity the $L_R$ vs $L_X$ relation seen in Fig. \ref{LRLX} 
for broad-line AGNs. This would account for the sharp decrease in the upper envelope of variability in Fig. \ref{Svar} from $\sim$0.3 to$\sim$0.03.
Of course, $\log L_X \la 42$ is the region of ``normal" and starburst galaxies, but dilution could bring low-luminosity AGNs in the same region.

\section{Conclusion}
The spectroscopic follow-up campaign of our sample of variability selected objects, with ESO NTT, was extremely successful 
since we obtained good quality spectra and most of the observed objects turned out to be interesting. 
The majority of the objects were clearly classified as BLAGNs, confirming the efficiency of variability surveys in detecting QSOs.
Some of the BLAGNs among our variable candidates (5/47) were missed by the X-ray surveys in the Chandra field, despite its depth. 
Some BLAGNs in our sample (9/48) have extended images (S$_{EIS}<$0.85) and would have been missed by the standard colour selection technique, 
which is restricted to point-like sources. This was in fact the original motivation of the search for ``variable galaxies" proposed 
by \citet{bers98}, in the field of SA 57.

Subsequent work by \citet{sara03,sara06} and \citet{kles07}, thanks to the high spatial resolution of HST images, allowed these authors to detect, 
through variability, active nuclei of very low luminosities, down to $M_B \approx -15$, comparable to those detectable in Seyfert galaxies 
closer to us. Some of these LLAGNs were even undetected in X-rays. The XMM survey of the SA 57 field \citep{trev07} and the follow-up 
spectroscopy \citep{trev08b} of a sample of variability- or X-ray-selected AGN candidates, which is still in progress (Zitelli et al. in preparation), 
suggested a complex scenario: while most high-luminosity ($L_X(2-10 keV) \ga 10^{43}$ erg s$^{-1}$), variability-selected AGNs are BLAGNs, below 
$L_X(2-10 keV) \approx 10^{42}$ erg s$^{-1}$ there is a spread in the $L_R-L_X$ plane. Although this spread is partly due to a difference in 
the dilution caused by the host galaxy, the above evidence raises at least two questions: i) what is the fraction of low X/O objects and its importance in 
understanding the global cosmological evolution of LLAGNs; ii) what is the physical origin of the low X-ray luminosity as compared with the emission in the 
optical band. According to \citet{gibs08}, the non-simultaneity of X-ray and optical observations may artificially increase the apparent scatter of the X/O and X/UV distribution, 
further complicating the comparison of the LF evolution as deduced from X-ray or optical observations. In any case, they conclude that the intrinsic spread of the 
X/O ratio is significantly lower than previous estimates in the literature. This conclusion puts constraints on the physical relation between the UV emitting accretion 
disk and the X-ray emitting corona. While the analysis of simultaneous observations, e.g. with XMM-Newton Optical Monitor (OM) or the SWIFT Ultraviolet Optical Telescope (UVOT) 
is obviously the main way to measure ``true" X/O an X/UV distributions, we stress that the discovery of X-ray undetected AGNs through optical variability and the study 
of their properties can provide precious information about the nature of X-ray weak AGNs.

Our previous results, on the sample of AGN candidates in the AXAF field (paper I) and in SA 57 \citep{trev08b}, confirm the existence of a population of NELGs, whose variability 
indicates the co-existence of starburst and nuclear activity. The present study, adding new follow-up spectroscopy of the AXAF sample, provides further support to this view.
In fact, 26\% of our variable candidates with extended image structure showed only narrow emission lines (NELG). For the 4 sources that we 
have been able to plot on the diagnostic diagram (see Fig. 6) we have clear evidence that at least 2 are intermediate kind of sources, where star 
formation co-exists with an active nucleus. For the other two sources, although in the diagnostic diagram they are in the area occupied by the star forming galaxies, 
the presence of variability is a strong indication for the presence of a faint active nucleus since star forming galaxies are not supposed to be variable. According 
to \citet{maoz05}, luminosity changes of stars cannot give such variability amplitudes. The nature of these sources is very interesting and should be further investigated. 
The most plausible scenario is the presence of a low-luminosity AGN and this makes optical variability extremely useful to pick up such weak active nuclei.
 
In the framework of the Subaru/XMM-Newton Deep Survey (SXDS) \citet{moro08b,moro08a} created a sample of AGN candidates, 
in $\approx 0.56$deg$^2$ of sky, and split it in X-ray-detected optically non-variable AGNs (XAs, 238 objects/36 with redshift), 
X-ray-detected optically variable AGNs (XVAs, 89 objects/35 with redshift), and X-ray-undetected optically variable AGNs (VAs, 112 objects/9 with redshift). 
The VAs are split, in turn, into 2 classes according to their variability properties, and a possible explanation in terms of Eddington ratio is suggested. 
At the moment only $\approx 20\%$ of the sample has been observed spectroscopically. Thus, despite the somewhat smaller area ($0.25 deg^2$) and 
optical depth ($V \simeq 24$) of our survey, our analysis already provides an important complement to the SXDS since, covering the CDFS and ECDFS 
fields, allowed us: i) to make use of several optical spectra already obtained by other authors, which have been complemented by the present follow-up 
campaign; ii) to push the analysis to the deeper X-ray flux limits reached by the Chandra observations in this field. Counting only the sources included in the fields 
with X-ray data and optical spectroscopy, we have 13 X-ray detected optically non variable BLAGN, 42 X-ray detected optically variable BLAGN and 5 X-ray undetected 
optically variable AGNs. The fact that our X-ray undetected sources are fewer than in the SXDS can be explained by the fact that the X-ray surveys 
in our field are a lot deeper (1 Ms for the CDFS in comparison to 100ks for the deep SXDS exposure).

An important result of this campaign was the identification of the source ID 26, which after our spectroscopic follow-up turned out to be a BLAGN but it showed 
no emission in the 1 Ms hard X-ray band and it was classified as galaxy by COMBO-17. The same characteristics are attributed to the source ID 125. This means that 4\% of 
our BLAGN were not detected by the traditional AGN selection techniques. One difference between the two sources is that the ID 125 was detected in the soft X-ray band, 
while ID 26 was not detected in any X-ray band and we have calculated the X-ray upper limit using the ECDFS intensity maps. This means that ID 26 is an AGN practically 
invisible to the traditional selection techniques and its was discovered solely on the base of its variability. We expect to find more of these sources as we go to fainter 
and more diffuse sources.

In Table 3 we give a detailed catalogue with the number of variable sources, X-ray detected and not, that we have identified. We also report, for each class of objects, 
the number of sources that lie in the field but, according to our study, do not show significant variability. These non variable sources are the ones represented by 
small dots in Fig. 9 and 10. They correspond to the sources with optical spectroscopy presented by \citet{szok04} and \citet{trei08}, for which we have a variability 
measurement (for details see paper I). We include also the numbers of type 2 AGNs from the literature, even though so far we have not identified 
such sources among our variable candidates. We note that 72\% of the total number of BLAGNs are variable, according to our selection.
   
Still, 45\% of our variability selected objects remain without optical spectroscopy, in part because previous spectroscopic studies were concentrated in the CDFS, while 
our sample covers the larger ECDFS, and in part because of the faint flux limit of the sample which requires 8-m class telescopes to obtain reliable optical spectra. 
This means that we still do not have enough data to study the luminosity function of our sample and put constraints on its evolution.
Since variability seems, in many cases, the sole mean to discover LLAGNs, we consider mandatory the completion of the follow-up spectroscopy for our 
sample which will provide a good statistical sample. Our result was made possible by the potential synergy between supernova searches and AGN detection. 
The full set of STRESS data are available and the creation of a catalogue of variable objects, with the characteristics of the present one,
can be easily produced. The ESSENCE supernova survey has also provided the largest deep sample of objects so far and its analysis is in progress 
(Boutsia et al., in preparation). Deep observations of those fields in the X-ray band would provide a unique contribution to the understanding of the complex, 
interesting LLAGN population and its evolution.

\begin{acknowledgements}
We thank P. Tozzi for providing us with the upper limits in the hard X-ray band (2-8keV), based on the ECDFS data. 
We thank R. Fosbury for advise in spectral classification and V. Mainieri for useful discussion of the X-ray data. K.B. 
acknowledges support by ESO studentship program 2006-2008. This work was partly supported by MIUR under grant PRIN 2006/025203. 
This research was supported by the DFG Cluster of Excellence ``The Origin and Structure of the Universe" (www.universe-cluster.de). 
This research has made use of the NASA/IPAC Extragalactic Database (NED) which is operated by the Jet Propulsion Laboratory, 
California Institute of Technology, under contract with the National Aeronautics and Space Administration.
\end{acknowledgements}

\input{1092T2T3.tex}

\begin{figure*}
\centering
\includegraphics[width=3cm]{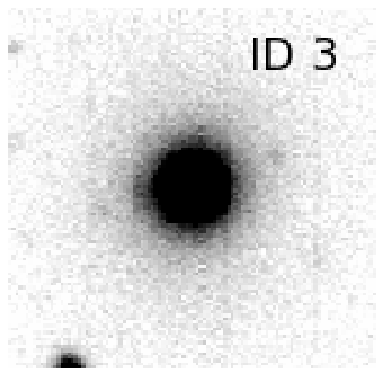}
\includegraphics[width=3cm]{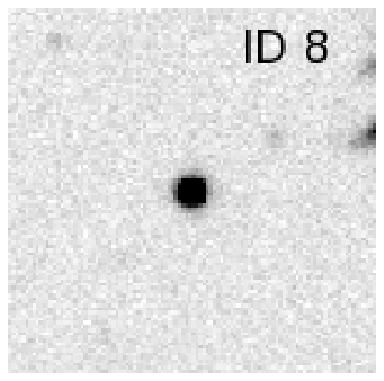}
\includegraphics[width=3cm]{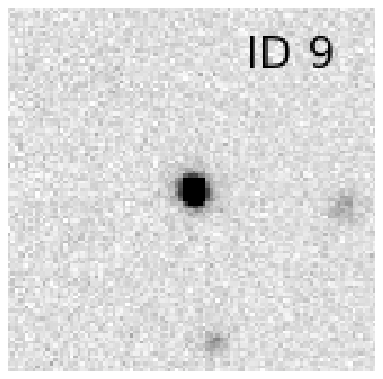}
\includegraphics[width=3cm]{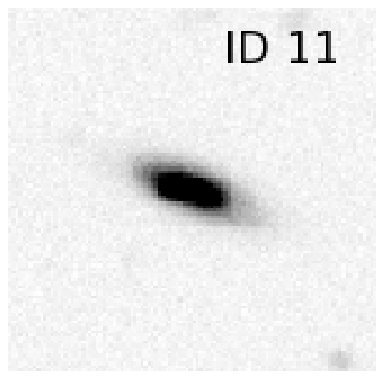}
\includegraphics[width=3cm]{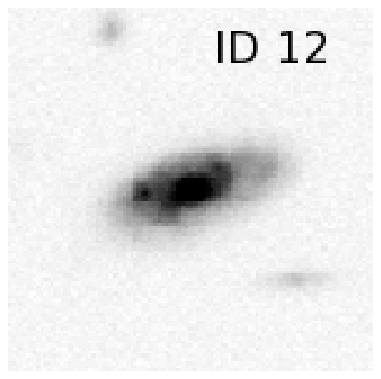}\\
\includegraphics[width=3cm]{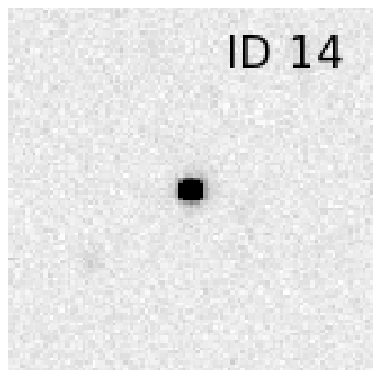}
\includegraphics[width=3cm]{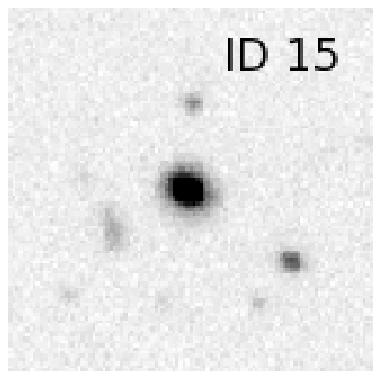}
\includegraphics[width=3cm]{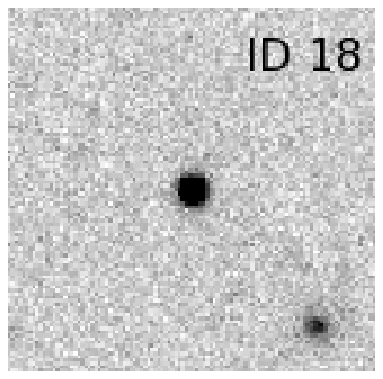}
\includegraphics[width=3cm]{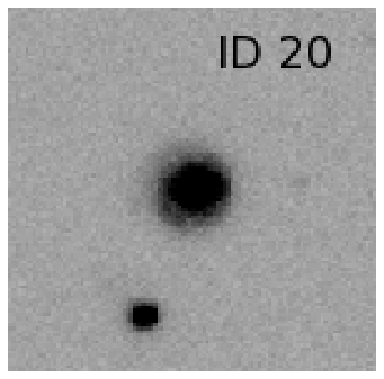}
\includegraphics[width=3cm]{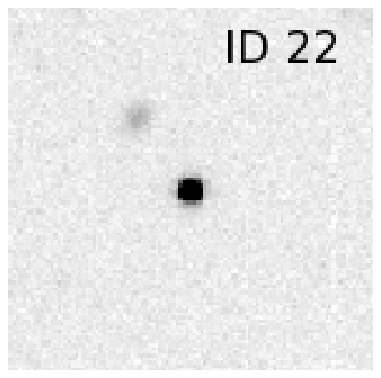}\\
\includegraphics[width=3cm]{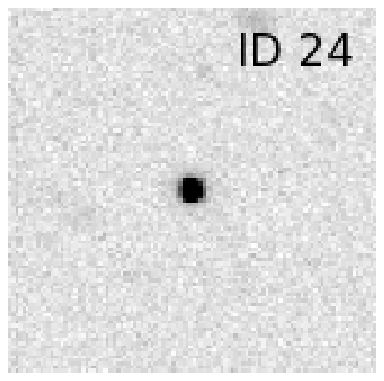}
\includegraphics[width=3cm]{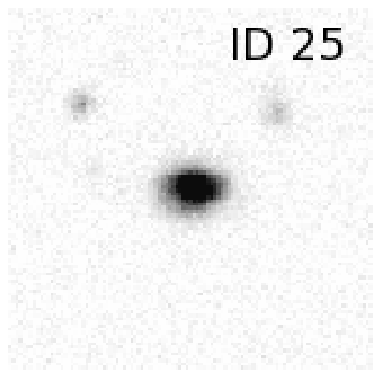}
\includegraphics[width=3cm]{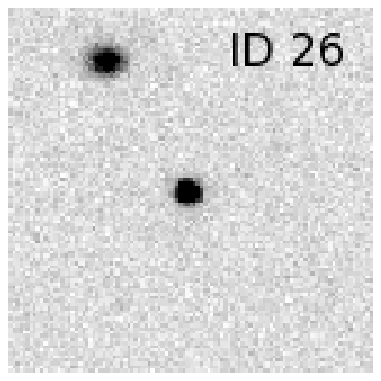}
\includegraphics[width=3cm]{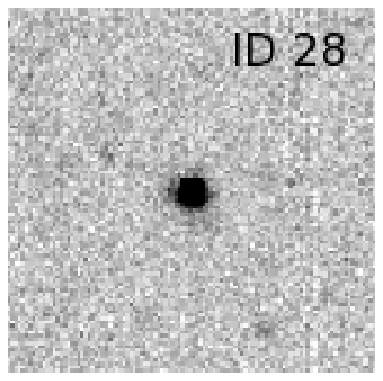}
\includegraphics[width=3cm]{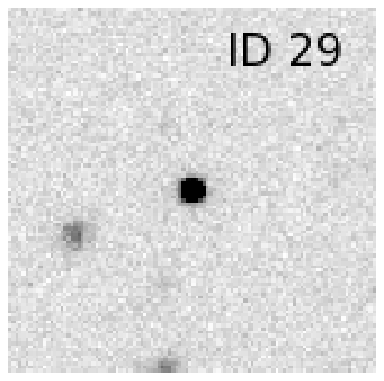}\\
\includegraphics[width=3cm]{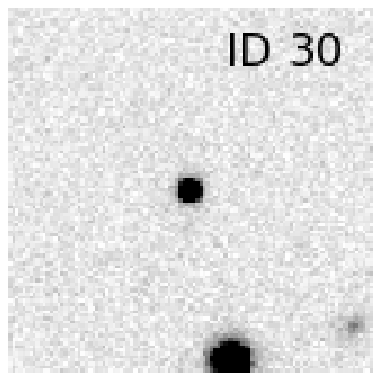}
\includegraphics[width=3cm]{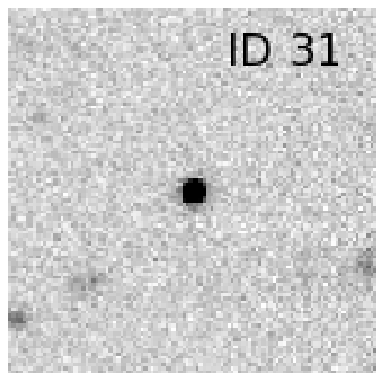}
\includegraphics[width=3cm]{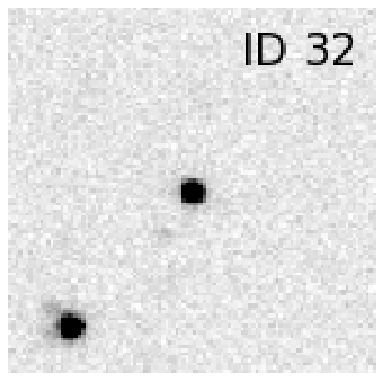}
\includegraphics[width=3cm]{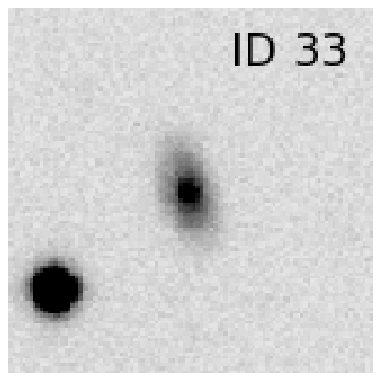}
\includegraphics[width=3cm]{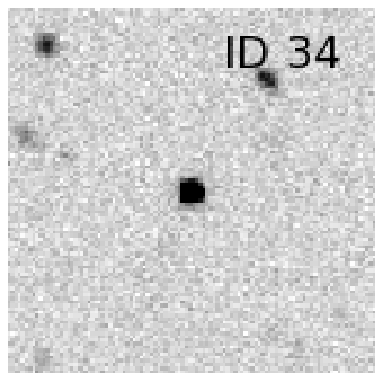}\\
\includegraphics[width=3cm]{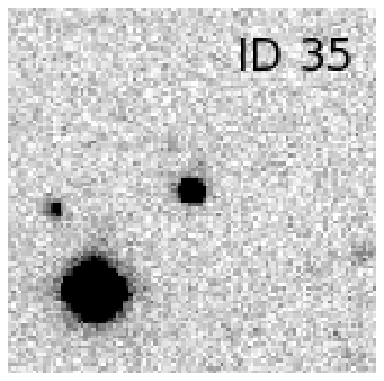}
\includegraphics[width=3cm]{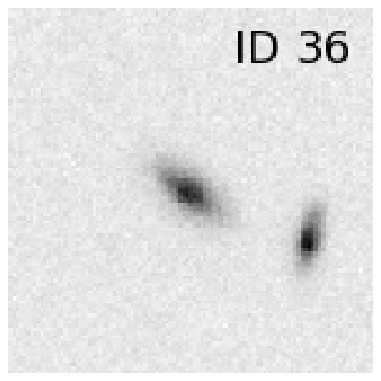}
\includegraphics[width=3cm]{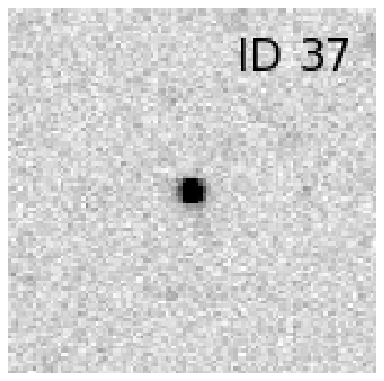}
\includegraphics[width=3cm]{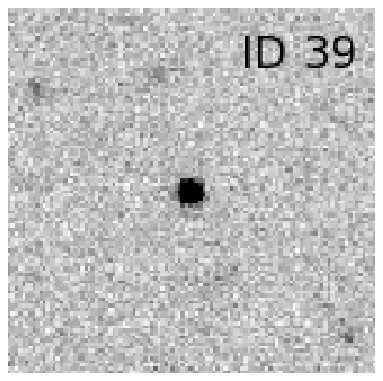}
\includegraphics[width=3cm]{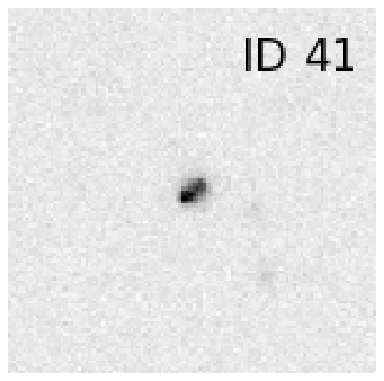}\\
\includegraphics[width=3cm]{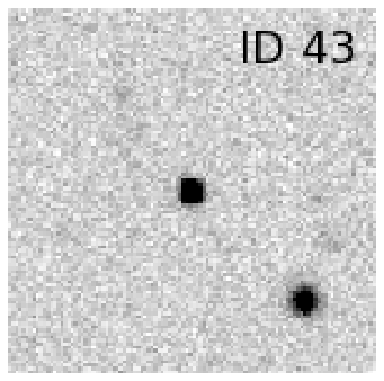}
\includegraphics[width=3cm]{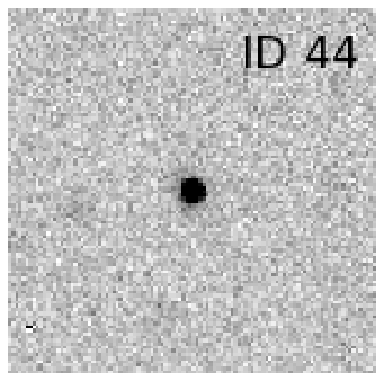}
\caption{\small Thumb-images ($29\times 29$ arcsec$^2$) of the sources in Table \ref{Tab1} obtained with WFI at ESO/MPI 2.2m with 900s exposure in the $V$ band. }
\label{thumbs}
\end{figure*}

\newpage
\bibliographystyle{aa}
\bibliography{biblio}{}

\appendix
\section{Comments on individual objects}
Here we present comments on the individual sources for which we have obtained optical spectra during this spectroscopic follow-up. 
 
\subsection{Objects with $\emph{z}$ $>$ 2}

\begin{itemize}
\item{\bf{ID 8}}: This source has the highest redshift of $\emph{z}_{em}$=3.16 in our sample. It is outside the area covered by EIS 
and COMBO-17, so the only information we have is our photometry and variability measurement. 
It shows a relatively low, yet significant variability of $\sigma$*=6.5. In the spectrum there are several broad foreground 
absorption lines, including a damped Ly$\alpha$ system at $\emph{z}_{abs}$=2.74. 
\item{\bf{ID 18}}: Here we find $\emph{z}_{em}$=2.60. This object has been classified as QSO by COMBO-17 and shows 
X-ray emission. The logarithm of the X-ray luminosity (in erg$s^{-1}$ here and in the following) is logL$_{X}$=44.25, which is typical of an AGN,
and the variability is significant with $\sigma$*=10.0. We find absorption features 
which correspond to Ly$\alpha$, NV and CIV at a redshift $\emph{z}_{abs}$=2.60. 
\item{\bf{ID 24}}: The redshift of this source is $\emph{z}_{em}$=2.56. Again, it is a COMBO-17 QSO with X-ray emission 
(logL$_{X}$=44.33) and $\sigma$*=11.0.  
\item{\bf{ID 30}}: Here we determine a redshift of $\emph{z}_{em}$=2.57. This source has typical AGN emission in the X-ray 
band, logL$_{X}$=44.73, and a stellarity index SI=0.98 according to EIS. By COMBO-17, it has been classified as galaxy, 
which means that the galaxy light dominated in the 17 bands. However, the variability we detect is significant.
This is an example of the objects that are missed by the colour selection and can be 
retrieved through their variability. Absorption lines are detected in the spectrum at a redshift, $\emph{z}_{abs}$=2.56, which 
agrees with the emission. 
\item{\bf{ID 34}}: This source has a redshift of $\emph{z}_{em}$=2.16. It is a COMBO-17 QSO with X-ray emission, logL$_{X}$=44.14, 
and $\sigma$*=10.3. The Ly$\alpha$ line is partially visible, but at the other end of the spectrum we can detect the MgII$\lambda$2798$\AA$ line.
\end{itemize}

\subsection{Objects with 1.5 $<$ $\emph{z}$ $<$ 2.0}

\begin{itemize}
\item{\bf{ID 14}}: This source is classified as QSO by COMBO-17 and its luminosity in the X-rays is logL$_{X}$=44.86. 
The variability measurement is rather high with $\sigma$*=11.9. 
It has a redshift of $\emph{z}_{em}$=1.95, which agrees with the redshift determination by \cite{norr06}. This source has also 
been detected in radio according to \cite{kell08}.
\item{\bf{ID 31}}: This source is a typical broad-line AGN but it lies outside the area covered by COMBO-17 and the ECDFS, 
therefor we have no SED classification or X-ray emission information available. The variability measurement is $\sigma$*=9.0 
and the redshift determined by the emission lines is $\emph{z}_{em}$=1.96.
\item{\bf{ID 32}}: This source has a redshift $\emph{z}_{em}$=1.62. It is classified as QSO by COMBO-17, it has an X-ray 
luminosity of logL$_{X}$=43.80 and shows high variability with $\sigma$*=18.6. 
\item{\bf{ID 35}}: Also this source is at redshift $\emph{z}_{em}$=1.62. It has emission in the X-rays, is a COMBO-17 QSO 
and its variability measurement is $\sigma$*=7.4.
\end{itemize}

\subsection{Objects with 1.0 $<$ $\emph{z}$ $<$ 1.5}

\begin{itemize}
\item{\bf{ID 26}}: This is one of the most interesting objects of this spectroscopic campaign. Even though the S/N is not 
particularly high, we securely detect CIII]$\lambda$1909$\AA$ and MgII$\lambda$2798$\AA$ at a redshift $\emph{z}_{em}$=1.33. 
For this object, even though it is well within the field of the ECDFS survey, no X-ray emission has been detected. 
Using the ECDFS intensity maps we estimate an upper limit in the hard X-ray band luminosity of logL$_{X}$$<$43.21. 
The luminosity in the R band is logL$_{R}$=44.45. This results to an X-ray-to-optical 
luminosity ratio of log($\emph{X/O}$)$<$-1.25, which is rather low for an AGN. Yet, its variability is high 
with $\sigma$*=13.6. Moreover, this object has been classified as galaxy by the COMBO-17 
survey, which means that the AGN is swamped by the galaxy light. This represents exactly the 
kind of objects, that are lost by other selection methods and can be retrieved by their variability.
\item{\bf{ID 28}}: This source has a redshift of $\emph{z}_{em}$=1.23. Even though it shows X-ray emission (logL$_{X}$=43.96) 
it has been classified as galaxy by COMBO-17. It has an X-ray-to-optical luminosity ratio log($\emph{X/O}$)=-0.38, which 
is consistent with its AGN nature. It has significant variability of $\sigma$*=7.3.
\item{\bf{ID 39}}: The redshift of this source is $\emph{z}_{em}$=1.15 and it is outside the area covered by the other 
catalogues, thus the only information we have is our variability measurement and magnitude. It exhibits a large variability with 
a very high significance, $\sigma$*=24.7. The broad line at around 8200$\AA$ is, most probably, second order contamination of the CIII] emission line.
\item{\bf{ID 44}}: The S/N of this spectrum is not very high, but we can clearly distinguish the two broad emission lines and 
we claim a secure redshift determination of $\emph{z}_{em}$=1.37, which agrees with the redshift reported by \cite{norr06}. 
This is another case of galaxy SED classification with X-ray emission typical of AGN (logL$_{X}$=44.52 and log($\emph{X/O}$)=0.31). 
The variability measurement is marginally significant with $\sigma$*=3.3. It also shows emission in the radio according to \cite{kell08}.
\end{itemize}

\subsection{Objects with 0.6 $<$ $\emph{z}$ $<$ 1.0}

\begin{itemize}
\item{\bf{ID 22}}: The redshift determination is based on MgII$\lambda$2798$\AA$ and the forbidden lines [SII]$\lambda$4069$\AA$ 
and [OIII]$\lambda$5007$\AA$. Broad H$\beta$ is marginally visible. We derive $\emph{z}_{em}$=0.84. This source has a high 
variability measurement of $\sigma$*=19.4, it has been classified as QSO by COMBO-17 and its emission in the X-ray band is 
typical of AGN (logL$_{X}$=44.17).
\item{\bf{ID 29}}: Even though the MgII$\lambda$2798$\AA$ line is not very prominent, the redshift $\emph{z}_{em}$=0.77 is 
also supported by the presence of [OII]$\lambda$3727$\AA$, [NeIII]$\lambda$3869$\AA$, H$\beta$ and the 
[OIII]$\lambda\lambda$4959,5007$\AA\AA$ doublet. This object is outside the fields of the other surveys and it is 
considerably variable with $\sigma$*=16.6.
\item{\bf{ID 37}}: This source has a redshift of $\emph{z}_{em}$=0.76. Apart from the broad MgII$\lambda$2798$\AA$ and H$\beta$ lines, 
there is a very prominent [OII]$\lambda$3727$\AA$ line and also [NeIII]$\lambda$3869$\AA$ and [OIII]$\lambda$5007$\AA$ are detected. 
It is a broad-line AGN with QSO SED and X-ray emission of logL$_{X}$=43.50.
\item{\bf{ID 43}}: This is another example of BLAGN, which was not selected as such by the COMBO-17 survey. Its variability 
measurement is high, $\sigma$*=20.9, it has X-ray emission typical of AGN (logL$_{X}$=43.05) and is at a redshift of $\emph{z}_{em}$=0.68. 
A wide range of emission lines are detected, including both broad lines (MgII$\lambda$2798$\AA$, H$\beta$), as well as forbidden 
high ionisation lines like [NeV]$\lambda$3426$\AA$, [NeIII]$\lambda$3869$\AA$ and [OIII]$\lambda\lambda$4959,5007$\AA\AA$.
\end{itemize}

\subsection{Objects with narrow emission lines (NELG)}

\begin{itemize}
\item{\bf{ID 11}}: A COMBO-17 galaxy with variability measurement $\sigma$*=4.0. A wide range of emission lines is detected 
and the redshift is $\emph{z}_{em}$=0.09, which agrees with the results of \cite{norr06}. The line ratios for this source are 
log([NII]/H$\alpha$)=-0.34 and log([OIII]/H$\beta$)=-0.09, which places it in the area of the composite objects in the diagnostic 
diagram \citep{kewl06}. The [OI]$\lambda$6300$\AA$ line is not as prominent as expected for AGN, but this could be due to the 
fact that it is right next to a sky absorption line. As mentioned above, this candidate has good possibilities of hosting a 
faint active nucleus. In our study, this object has $V=19.39$ which is quite different from the V$_{EIS}$=17.78. This happens 
with all the bright sources with extended image structure and it is explained by the fact that our magnitudes are calculated for 
a fixed aperture, instead in the EIS catalogue, the total magnitude of the source is reported. This source is detected in radio 
by \cite{kell08} but the emission is low, compatible with a normal galaxy. We calculated the upper limit of the X-ray luminosity in 
the ECDFS field and it corresponds to logL$_{X}<$40.86.
\item{\bf{ID 12}}: This candidate has $\emph{z}_{em}$=0.10 based on a wide range of emission lines and this value 
agrees with the redshift determination by \cite{way05}. It is classified as galaxy by COMBO-17 and the variability 
measurement is $\sigma$*=3.7. Since all the necessary lines are detected, we derive line ratios. 
We have log([NII]/H$\alpha$)=-0.31 and log([OIII]/H$\beta$)=0.12 and these values are consistent again with the composite object 
locus in the diagnostic diagram. This object is within the CDFS area and we have calculated a 3$\sigma$ upper limit for its flux 
in the 2-8keV band that corresponds to a luminosity of logL$_{X}<$40.37. 
\item{\bf{ID 15}}: The redshift of this source is $\emph{z}_{em}$=0.21 and it has a variability significance of $\sigma$*=5.0. 
We find log([NII]/H$\alpha$)=-0.41 and log([OIII]/H$\beta$)=-0.31 which suggest a star-forming galaxy. Yet, considering the 
significant variability we cannot exclude the presence of a faint AGN and for this reason we classify it as NELG.
\item{\bf{ID 20}}: This candidate has the largest variability measurement in this class of objects with $\sigma$*=6.8. 
The emission lines indicate a redshift of $\emph{z}_{em}$=0.24. Apart from the 
H$\alpha\lambda$6563$\AA$, [NII]$\lambda$6584$\AA$ doublet and the [OII]$\lambda$3727$\AA$ forbidden line, no 
other lines are detected so we cannot calculate line ratios. The [SII]$\lambda\lambda$6717,6731$\AA\AA$ 
doublet is visible but not very prominent. This source is not in the field of the two X-ray surveys, thus we have no upper limit 
for the X-ray emission.
\item{\bf{ID 25}}: This object is in the ECDFS field and the upper limit of the X-ray luminosity corresponds to logL$_{X}<$40.73. 
We derive a redshift of $\emph{z}_{em}$=0.10 and the variability is $\sigma$*=3.5. The H$\beta$ line is not visible, thus we can derive
only a lower limit on the [OIII]/H$\beta$ ratio. However, the ratio log([NII]/H$\alpha$)=-0.58 places this object in an area where 
it is most likely a star-forming galaxy. 
\item{\bf{ID 33}}: Here the only lines clearly detected are H$\alpha$, the [NII]$\lambda$6584$\AA$ doublet and the 
[SII]$\lambda\lambda$6717,6731$\AA\AA$ doublet at $\emph{z}_{em}$=0.15. We see that the wings of the narrow 
emission lines are slightly broadened. This could suggest a weak AGN. It is worth mentioning that this candidate 
is in the region of the CDFS survey, but it shows no X-ray emission and the upper limit we obtain is logL$_{X}<$40.58.
\item{\bf{ID 36}}: The redshift of this candidate is $\emph{z}_{em}$=0.15 based on H$\alpha$, [SII]$\lambda$6717$\AA$ 
and [OII]$\lambda$3727$\AA$. The [NII]$\lambda$6584$\AA$ line has been suppressed by the sky subtraction. The significance 
of the variability is $\sigma$*=3.6. The [OIII]$\lambda$5007$\AA$ is marginally detected. Also this source is outside the X-ray survey fields.
\end{itemize}

\subsection{Other}

\begin{itemize}
\item{\bf{ID 3}}: This source has a spectrum typical of normal galaxies with $\emph{z}_{abs}$=0.13 and it is 
within the ECDFS field. The redshift determination agrees with the value given by \cite{way05}. It exhibits X-ray emission 
in the soft band and an upper limit has been determined for its emission in the 2-8keV band. This upper limit for the X-ray 
luminosity in the hard band, logL$_{X}<$40.74, is lower than the typical AGN values (42$<$logL$_{X}<$45) and everything 
seems to suggest a normal galaxy. This source has been detected in radio \citep{kell08} and its emission also 
suggests a normal galaxy. However, normal galaxies are not supposed to vary and this candidate has significant 
variability measurement with $\sigma$*=3.8.
\item{\bf{ID 9}}: This candidate is a K-star. Its variability measurement is right on the threshold with $\sigma$*=3.0.
\item{\bf{ID 41}}: This is an M-star with a variability significance $\sigma$*=5.8.
It has been characterised as Galaxy/Unclear by the COMBO-17 survey and in the EIS catalogue it has a SI=0.03 which indicates an extended 
image structure. This peculiarity can be explained by the fact that this object is in fact two blended sources (see thumb in Fig.\ref{thumbs}) although 
there is no indication that it is a physical binary. 
\end{itemize}

\end{document}

%% file: 1092T1.tex
\begin{table*}[htbp]
\centering
\caption{Catalogue of variable sources observed during our spectroscopic follow-up.}
\begin{tabular}{rcccccccl}
\hline\hline             
ID&RA&DEC&$V$&$V_{min}-V_{max}$&exp.time&airmass&z&S-class\\
\hline
  3&03:33:20.61&-27:49:10.1&18.47& 0.11 &900 &1.01&0.13&Gal  \\
  8&03:32:46.76&-28:08:46.7&19.21& 0.17 &1800&1.04&3.16&BLAGN\\
  9&03:33:38.86&-27:40:16.4&19.24& 0.11 &1800&1.11&-   &Star \\
 11&03:33:16.51&-27:50:39.5&19.39& 0.15 &900 &1.01&0.09&NELG \\
 12&03:32:45.95&-27:57:45.3&19.60& 0.15 &900 &1.27&0.10&NELG \\
 14&03:32:32.00&-28:03:09.9&19.86& 0.30 &900 &1.17&1.95&BLAGN\\
 15&03:31:54.65&-28:10:35.7&19.88& 0.11 &900 &1.10&0.21&NELG \\
 18&03:33:31.37&-27:56:34.2&20.01& 0.23 &900 &1.04&2.60&BLAGN\\
 20&03:32:11.84&-28:09:11.0&20.03& 0.13 &900 &1.10&0.24&NELG \\
 22&03:33:28.93&-27:56:41.1&20.05& 0.41 &900 &1.04&0.84&BLAGN\\
 24&03:33:09.71&-27:56:14.0&20.21& 0.27 &900 &1.27&2.56&BLAGN\\
 25&03:31:16.69&-27:43:29.6&20.28& 0.16 &1800&1.09&0.10&NELG \\
 26&03:32:34.57&-28:03:14.0&20.29& 0.29 &900 &1.17&1.33&BLAGN\\
 28&03:33:21.09&-27:39:11.8&20.48& 0.23 &1800&1.11&1.22&BLAGN\\
 29&03:32:37.29&-28:08:47.0&20.50& 0.37 &1800&1.04&0.77&BLAGN\\
 30&03:33:12.63&-27:55:51.6&20.50& 0.16 &1800&1.18&2.57&BLAGN\\
 31&03:31:11.38&-27:41:31.9&20.62& 0.27 &1800&1.09&1.96&BLAGN\\
 32&03:33:22.79&-27:55:23.8&20.66& 0.44 &2700&1.25&1.61&BLAGN\\
 33&03:32:53.90&-27:53:54.1&20.74& 0.12 &1800&1.18&0.15&NELG \\
 34&03:33:26.24&-27:58:29.7&20.79& 0.27 &2700&1.25&2.16&BLAGN\\
 35&03:32:20.30&-28:02:14.8&20.84& 0.18 &1800&1.09&1.62&BLAGN\\
 36&03:32:31.78&-28:07:10.4&20.84& 0.16 &1800&1.33&0.15&NELG \\
 37&03:33:29.22&-27:59:26.7&21.03& 0.22 &1800&1.21&0.76&BLAGN\\
 39&03:32:44.18&-28:10:28.5&21.07& 0.49 &1800&1.33&1.15&BLAGN\\
 41&03:31:58.13&-28:02:41.5&21.14& 0.17 &1800&1.09&-   &Star \\
 43&03:33:20.01&-27:59:12.4&21.22& 0.46 &1800&1.21&0.68&BLAGN\\
 44&03:31:15.04&-27:55:18.6&21.25& 0.14 &1800&1.42&1.37&BLAGN\\
\hline
\end{tabular}
\label{Tab1}
\end{table*}

%% file: 1092T2T3.tex
\pagestyle{empty}

\begin{sidewaystable*}
\hspace{-1cm}
\begin{minipage}[t][190mm]{\textwidth\scriptsize}
\caption{Catalogue of extragalactic variable objects in the AXAF field with spectroscopic redshift.}
\label{Tab2}
 \begin{tabular}{rccccrcccrrccccccrrcrl}
\hline\hline             
ID&$\alpha$(J2000)&$\delta$(J2000)&$V$&$\sigma$&$\sigma^{*}$&EIS-name&$V_{EIS}$&$R_{EIS}$&$(U-B)_{EIS}$&$(B-V)_{EIS}$&$S_{EIS}$&COMBO&class$^a$&$z$&Spclass$^b$&XID&$f_X$(2-8keV)&L$_{X}$&L$_{R}$&log(X/O)&notes\\
(1)&(2)&(3)&(4)&(5)&(6)&(7)&(8)&(9)&(10)&(11)&(12)&(13)&(14)&(15)&(16)&(17)&(18)&(19)&(20)&(21)&(22)\\
\hline
  3& 03:33:20.61& -27:49:10.1&18.47& 0.032&  3.8 & J033320.61-274910.3&   16.68 & 16.12  & 1.49  & 0.95  &  0.03 & 32802& G  &0.13  &Gal    &664& $<$1.42e-15  &  $<$40.74 &  43.66  &$<$-2.923&c,p  \\
  4& 03:32:35.09& -27:55:33.0&18.72& 0.047&  5.2 & J033235.09-275533.2&   16.77 & 16.21  & 1.32  & 0.49  &  0.03 & 18675& G  &0.038 &NELG   &247&  1.46e-15    &   39.63   &  42.50  &  -2.877 &d,m,n,p\\
  5& 03:32:29.99& -27:44:04.8&18.75& 0.028&  3.4 & J033230.00-274405.0&   17.50 & 17.10  & 1.05  & 0.40  &  0.03 & 42499& G  &0.076 &NELG   &392&  3.05e-15    &   40.57   &  42.77  &  -2.200 &d,m,p\\
  6& 03:32:27.01& -27:41:05.1&19.14& 0.077& 13.0 & J033227.02-274105.2&   19.12 & 19.12  & 0.42  & 0.01  &  0.98 & 48284& Q  &0.734 &BLAGN  &379&  6.58e-14    &   44.14   &  44.20  &  -0.058 &d    \\
  8& 03:32:46.76& -28:08:46.7&19.21& 0.055&  6.5 & J033246.76-280846.8& 	&	 &	 &	 &	 &	&    &3.16  &BLAGN  &	&	       &	   &	     &         &c    \\
 10& 03:32:08.66& -27:47:34.4&19.25& 0.059&  9.2 & J033208.68-274734.5&   19.08 & 18.76  & 0.71  &-0.01  &  0.98 & 34357& Q  &0.543 &BLAGN  &305&  7.05e-14    &   43.85   &  44.03  &  -0.173 &d,p  \\
 11& 03:33:16.51& -27:50:39.5&19.39& 0.044&  4.0 & J033316.51-275039.7&   17.78 & 17.27  & 1.19  & 0.55  &  0.07 & 28467& G  &0.09  &NELG   &	& $<$4.11e-15  &  $<$40.86 &  42.86  &$<$-2.002&c,l,m,p\\
 12& 03:32:45.95& -27:57:45.3&19.60& 0.043&  3.7 & J033245.97-275745.6&   17.26 & 16.89  & 1.16  & 0.41  &  0.03 & 14012& G/U&0.10  &NELG   &	& $<$1.06e-15  &  $<$40.37 &  43.11  &$<$-2.743&c,k,p\\
 14& 03:32:32.00& -28:03:09.9&19.86& 0.100& 11.9 & J033232.02-280310.0&   19.82 & 19.72  &-0.20  & 0.03  &  0.98 &  2006& Q  &1.95  &BLAGN  &398&  3.11e-14    &   44.86   &  45.01  &  -0.144 &c,p  \\
 15& 03:31:54.65& -28:10:35.7&19.88& 0.032&  5.0 & ---  	     &  	&	 &	 &	 &	 &	&    &0.21  &NELG   &	&	       &	   &	     &         &c    \\
 18& 03:33:31.37& -27:56:34.2&20.01& 0.089& 10.0 & J033331.39-275634.6&   20.07 & 19.90  & 0.57  & 0.08  &  0.98 & 15952& Q  &2.60  &BLAGN  &723&  3.71e-15    &   44.25   &  45.24  &  -0.994 &c    \\
 19& 03:32:26.50& -27:40:35.7&20.02& 0.120& 17.7 & J033226.51-274035.7&   20.07 & 19.94  & 0.36  &-0.09  &  0.97 & 49298& Q  &1.031 &BLAGN  &375&  6.16e-15    &   43.48   &  44.24  &  -0.758 &d    \\
 20& 03:32:11.84& -28:09:11.0&20.03& 0.040&  6.8 & ---  	     &  	&	 &	 &	 &	 &	&    &0.24  &NELG   &	&	       &	   &	     &         &c    \\
 22& 03:33:28.93& -27:56:41.1&20.05& 0.155& 19.4 & J033328.95-275641.4&   20.12 & 20.12  & 0.47  & 0.13  &  0.98 & 15731& Q  &0.84  &BLAGN  &712&  4.98e-14    &   44.17   &  43.94  &   0.222 &c    \\
 23& 03:32:16.20& -27:39:30.2&20.18& 0.102& 14.5 & J033216.21-273930.5&   20.20 & 19.82  & 0.38  & 0.16  &  0.97 & 51593& G/U&1.324 &BLAGN  &345&  7.69e-15    &   43.84   &  44.55  &  -0.711 &d    \\
 24& 03:33:09.71& -27:56:14.0&20.21& 0.096& 11.0 & J033309.72-275614.3&   20.11 & 20.05  & 0.40  & 0.04  &  0.98 & 16621& Q  &2.56  &BLAGN  &596&  4.70e-15    &   44.33   &  45.16  &  -0.832 &c    \\
 25& 03:31:16.69& -27:43:29.6&20.28& 0.048&  3.5 & J033116.70-274329.5&   19.07 & 18.78  & 0.91  & 0.36  &  0.03 & 43124& G  &0.10  &NELG   &	& $<$2.42e-15  &  $<$40.73 &  42.35  &$<$-1.629&c,l  \\
 26& 03:32:34.57& -28:03:14.0&20.29& 0.113& 13.6 & J033234.59-280314.1&   20.20 & 20.08  & 0.61  & 0.10  &  0.98 &  1821& G/U&1.33  &BLAGN  &	& $<$1.77e-15  &  $<$43.21 &  44.45  &$<$-1.245&c,l  \\
 27& 03:32:38.12& -27:39:44.9&20.48& 0.058&  6.0 & J033238.14-273945.0&   20.43 & 20.41  & 0.05  & 0.41  &  0.98 & 50997& Q  &0.837 &BLAGN  &417&  1.47e-14    &   43.63   &  43.82  &  -0.193 &d    \\
 28& 03:33:21.09& -27:39:11.8&20.48& 0.078&  7.3 & J033321.09-273912.1&   20.39 & 20.14  & 0.11  & 0.00  &  0.83 & 52280& G/U&1.22  &BLAGN  &670&  1.23e-14    &   43.96   &  44.34  &  -0.378 &c    \\
 29& 03:32:37.29& -28:08:47.0&20.50& 0.135& 16.6 & J033237.32-280847.3& 	&	 &	 &	 &	 &	&    &0.77  &BLAGN  &	&	       &	   &	     &         &c    \\
 30& 03:33:12.63& -27:55:51.6&20.50& 0.063&  6.1 & J033312.63-275551.9&   20.38 & 20.08  & 0.96  & 0.05  &  0.98 & 17446& G  &2.57  &BLAGN  &611&  1.17e-14    &   44.73   &  45.16  &  -0.425 &c    \\
 31& 03:31:11.38& -27:41:31.9&20.62& 0.102&  9.0 & J033111.39-274131.9&   20.62 & 20.57  &-0.66  & 0.34  &  0.98 &	&    &1.96  &BLAGN  &	&	       &	   &  44.67  &         &c    \\
 32& 03:33:22.79& -27:55:23.8&20.66& 0.150& 18.6 & J033322.80-275524.0&   20.63 & 20.58  &-0.10  & 0.41  &  0.98 & 18256& Q  &1.61  &BLAGN  &678&  4.25e-15    &   43.80   &  44.46  &  -0.664 &c    \\
 33& 03:32:53.90& -27:53:54.1&20.74& 0.039&  3.1 & J033253.90-275354.3&   19.13 & 18.75  & 0.96  & 0.53  &  0.03 & 21830& G  &0.15  &NELG   &	& $<$7.14e-16  &  $<$40.58 &  42.75  &$<$-2.172&c,k  \\
 34& 03:33:26.24& -27:58:29.7&20.79& 0.092& 10.3 & J033326.26-275830.0&   20.65 & 20.57  &-0.11  & 0.04  &  0.98 & 11941& Q  &2.16  &BLAGN  &700&  4.54e-15    &   44.14   &  44.78  &  -0.640 &c    \\
 35& 03:32:20.30& -28:02:14.8&20.84& 0.053&  7.4 & J033220.34-280214.8&   20.76 & 20.63  &-0.14  & 0.41  &  0.93 &  4050& Q  &1.62  &BLAGN  &358&  8.71e-15    &   44.11   &  44.45  &  -0.332 &c    \\
 36& 03:32:31.78& -28:07:10.4&20.84& 0.045&  3.6 & J033231.81-280710.6& 	&	 &	 &	 &	 &	&    &0.15  &NELG   &	&	       &	   &	     &         &c    \\
 37& 03:33:29.22& -27:59:26.7&21.03& 0.075&  7.7 & J033329.24-275927.1&   20.95 & 20.71  & 0.03  & 0.42  &  0.92 &  9954& Q/G&0.76  &BLAGN  &716&  1.37e-14    &   43.50   &  43.60  &  -0.104 &c    \\
 38& 03:32:39.09& -27:46:01.8&21.05& 0.051&  4.6 & J033239.10-274602.0&   21.01 & 20.81  &-0.18  & 0.28  &  0.98 & 37487& Q  &1.216 &BLAGN  &423&  7.09e-15    &   43.72   &  44.07  &  -0.351 &d    \\
 39& 03:32:44.18& -28:10:28.5&21.07& 0.193& 24.7 & ---  	     &  	&	 &	 &	 &	 &	&    &1.15  &BLAGN  &	&	       &	   &	     &         &c    \\
 40& 03:32:09.44& -27:48:06.8&21.10& 0.058&  6.7 & J033209.46-274806.9&   20.96 & 20.50  & 1.82  & 0.16  &  0.98 & 33069& Q  &2.810 &BLAGN  &309&  2.22e-15    &   44.10   &  45.08  &  -0.978 &d    \\
 43& 03:33:20.01& -27:59:12.4&21.22& 0.167& 20.9 & J033320.02-275912.7&   21.26 & 20.92  & 0.54  & 0.00  &  0.98 & 10418& G  &0.68  &BLAGN  &661&  6.46e-15    &   43.05   &  43.40  &  -0.345 &c    \\
 44& 03:31:15.04& -27:55:18.6&21.25& 0.047&  3.3 & J033115.05-275518.8&   21.21 & 20.77  &-0.55  & 0.53  &  0.91 & 18408& G  &1.37  &BLAGN  &  7&  3.35e-14    &   44.52   &  44.21  &   0.310 &c,p  \\
 47& 03:32:29.98& -27:45:29.9&21.42& 0.169& 18.4 & J033229.99-274530.1&   21.52 & 21.24  & 0.10  &-0.11  &  0.98 & 38551& Q  &1.218 &BLAGN  &391&  1.08e-14    &   43.90   &  43.90  &   0.004 &d    \\
 48& 03:31:35.43& -28:03:15.8&21.43& 0.051&  3.1 & J033135.44-280315.8&   21.39 & 20.95  & 0.40  & 0.34  &  0.90 &  1647& G/U&1.348 &BLAGN  & 94&  4.48e-15    &   43.63   &  44.12  &  -0.492 &s    \\ 
 49& 03:32:59.86& -27:47:48.2&21.45& 0.205& 22.8 & J033259.85-274748.4&   21.66 & 21.92  & 0.20  & 0.43  &  0.98 & 33644& Q  &2.579 &BLAGN  &526&  9.59e-15    &   44.65   &  44.42  &   0.226 &d    \\
 50& 03:33:10.63& -27:57:48.5&21.52& 0.097& 10.6 & J033310.64-275748.8&   21.45 & 21.14  & 0.21  & 0.02  &  0.96 & 13332& Q  &1.598 &BLAGN  &601&  6.41e-15    &   43.97   &  44.23  &  -0.261 &s    \\ 
 55& 03:31:27.79& -28:00:51.0&21.75& 0.157& 13.0 & J033127.80-280051.2&   21.76 & 21.53  &-0.24  & 0.02  &  0.98 &  6817& Q  &1.950 &BLAGN  & 54&  5.00e-15    &   44.07   &  44.28  &  -0.213 &s    \\ 
 56& 03:33:32.75& -27:49:07.8&21.76& 0.236& 18.9 & J033332.77-274908.0&   22.09 & 21.64  & 0.64  &-0.50  &  0.98 & 31085& Q  &3.031 &BLAGN  &728&  6.63e-15    &   44.66   &  44.70  &  -0.046 &s    \\ 
 60& 03:32:32.28& -28:03:28.3&21.88& 0.168& 14.1 & J033232.30-280328.4&   21.82 & 21.41  &-0.04  & 0.07  &  0.92 &  1257& Q  &1.220 &BLAGN  &400&  1.88e-14    &   44.14   &  43.83  &   0.314 &s    \\ 
 61& 03:31:36.25& -28:01:49.7&21.92& 0.466& 36.3 & J033136.25-280149.8&   22.55 & 22.27  & 0.40  &-0.75  &  0.97 &  4809& Q  &1.953 &BLAGN  &100&  5.71e-15    &   44.13   &  43.99  &   0.141 &s    \\ 
 62& 03:32:59.07& -27:43:39.5&21.92& 0.062&  3.3 & J033259.06-274339.8&   21.48 & 20.99  & 0.31  & 0.47  &  0.04 & 42601& Q/G&0.733 &BLAGN  &516&  6.13e-15    &   43.11   &  43.45  &  -0.340 &d    \\
 70& 03:32:30.19& -28:00:19.9&22.05& 0.138&  9.7 & J033230.21-280020.0&   21.86 & 21.93  & 0.56  & 0.22  &  0.90 &  7902& Q  &2.305 &BLAGN  &393&  6.70e-15    &   44.38   &  44.30  &   0.074 &s    \\ 
 72& 03:31:35.78& -27:51:34.9&22.11& 0.161&  8.3 & J033135.79-275134.9&   21.95 & 21.65  & 0.79  &-0.33  &  0.98 & 25884& Q  &1.624 &BLAGN  & 96&  1.49e-14    &   44.35   &  44.04  &   0.309 &s    \\ 
 74& 03:31:56.86& -28:01:48.7&22.27& 0.276& 18.5 & J033156.88-280149.0&   22.37 & 21.94  & 0.36  &-0.16  &  0.83 &  4995& Q  &1.380 &BLAGN  &235& $<$1.34e-15  &  $<$43.13 &  43.75  &$<$-0.621&s    \\ 
 75& 03:32:17.14& -27:43:03.3&22.31& 0.091&  4.7 & J033217.15-274303.5&   22.16 & 21.14  & 0.63  & 0.37  &  0.85 & 43863& G  &0.569 &BLAGN  &348&  6.29e-15    &   42.85   &  43.12  &  -0.270 &d,p  \\
 76& 03:32:00.37& -27:43:19.7&22.33& 0.192& 12.0 & J033200.37-274319.9&   22.25 & 22.13  &-0.23  & 0.74  &  0.98 & 43151& Q  &1.037 &BLAGN  &250&  6.09e-15    &   43.48   &  43.37  &   0.113 &d    \\
 77& 03:31:51.78& -28:00:25.6&22.37& 0.109&  4.7 & J033151.80-280025.9&   22.38 & 22.38  & 0.30  & 0.20  &  0.98 &  7671& Q  &2.436 &BLAGN  &199&  2.25e-15    &   43.96   &  44.18  &  -0.220 &s    \\ 
 78& 03:31:50.95& -27:41:15.9&22.44& 0.160&  8.6 & J033150.97-274116.1&   22.67 & 22.21  & 0.44  & 0.08  &  0.95 & 47615& Q  &0.253 &NELG   &192&  6.32e-15    &   42.03   &  41.87  &   0.160 &d    \\
 81& 03:33:00.78& -27:55:20.7&22.52& 0.529& 28.5 & J033300.73-275520.6&   22.25 & 22.34  &-0.27  & 0.75  &  0.37 & 18324& Q/G&2.005 &BLAGN  &532&  1.42e-14    &   44.55   &  43.99  &   0.565 &s    \\ 
 83& 03:31:49.54& -27:43:19.4&22.71& 0.178&  7.6 & J033149.55-274319.6&   22.40 & 21.88  & 0.23  & 0.53  &  0.97 & 43170& G  &1.320 &BLAGN  &184&  7.31e-15    &   43.82   &  43.73  &   0.092 &e    \\
 84& 03:32:10.91& -27:44:15.0&22.75& 0.323& 15.2 & J033210.92-274415.2&   22.90 & 22.42  & 0.31  & 0.14  &  0.88 & 41159& Q  &1.600 &BLAGN  &316&  1.27e-14    &   44.26   &  43.72  &   0.548 &d,p  \\
 87& 03:33:22.85& -28:03:13.0&22.85& 0.238&  8.6 & J033322.87-280313.2&   22.71 & 22.32  & 0.55  &-0.11  &  0.71 &  1731& Q  &2.920 &BLAGN  &679& $<$1.58e-15  &  $<$44.00 &  44.39  &$<$-0.397&s    \\ 
 88& 03:32:38.87& -27:59:18.7&22.86& 0.146&  4.0 & J033238.88-275918.9&   22.58 & 22.04  & 0.11  & 0.64  &  0.81 & 10151& G  &0.651 &Gal    &419&  9.43e-15    &   43.17   &  42.91  &   0.266 &g,p  \\
 91& 03:33:16.08& -28:01:31.3&22.92& 0.135&  3.4 & J033316.10-280131.5&   22.52 & 22.21  &-0.48  & 0.34  &  0.60 &  5498& Q/G&2.072 &BLAGN  &631&  3.94e-15    &   44.03   &  44.08  &  -0.044 &s    \\ 
 94& 03:33:10.19& -27:48:42.0&22.96& 0.208&  5.1 & J033310.19-274842.3&   23.22 & 22.77  & 0.55  & 0.50  &  0.87 & 31898& Q/G&1.029 &BLAGN  &597& $<$1.43e-15  &  $<$42.84 &  43.11  &$<$-0.264&h,p  \\
 96& 03:32:43.24& -27:49:14.1&22.97& 0.332& 13.0 & J033243.25-274914.4&   22.70 & 22.65  & 0.01  &-0.17  &  0.95 & 30792& Q  &1.920 &BLAGN  &441&  2.63e-15    &   43.78   &  43.82  &  -0.044 &d    \\
 98& 03:32:41.86& -27:52:02.5&23.06& 0.143&  3.6 & J033241.86-275202.6&   23.14 & 22.52  &	 &	 &  0.91 & 25042& Q  &3.592 &BLAGN  &435&  2.86e-15    &   44.47   &  44.53  &  -0.060 &d    \\
\hline
\end{tabular}
\end{minipage}
\end{sidewaystable*}

\begin{sidewaystable*}
\hspace{-1cm}
\begin{minipage}[t][190mm]{\textwidth\scriptsize}
 \begin{tabular}{rccccrcccrrccccccrrcrl}
\hline\hline             
ID&$\alpha$(J2000)&$\delta$(J2000)&$V$&$\sigma$&$\sigma^{*}$&EIS-name&$V_{EIS}$&$R_{EIS}$&$(U-B)_{EIS}$&$(B-V)_{EIS}$&$S_{EIS}$&COMBO&class$^a$&$z$&Spclass$^b$&XID&$f_X$(2-8keV)&L$_{X}$&L$_{R}$&log(X/O)&notes\\
(1)&(2)&(3)&(4)&(5)&(6)&(7)&(8)&(9)&(10)&(11)&(12)&(13)&(14)&(15)&(16)&(17)&(18)&(19)&(20)&(21)&(22)\\
\hline
103& 03:32:32.50& -27:39:02.4&23.25& 0.226&  6.3 & J033232.52-273902.6&   23.33 & 22.89  & 1.33  & 0.75  &  0.07 & 52474& G  &0.572 &Gal    &	& $<$1.67e-15  &  $<$42.28 &  42.43  &$<$-0.145&g,k  \\
115& 03:32:01.58& -27:43:27.0&23.66& 0.335&  8.5 & J033201.59-274327.2&   23.34 & 23.06  & 1.28  & 0.24  &  0.97 & 42882& Q  &2.726 &BLAGN  &260&  4.88e-15    &   44.41   &  44.03  &   0.387 &d    \\
120& 03:33:14.85& -27:57:49.1&23.79& 0.204&  3.1 & J033314.86-275749.3&   23.96 & 23.48  & 1.65  &-0.26  &  0.87 & 13244& Q  &2.796 &BLAGN  &625&  1.05e-15    &   43.77   &  43.88  &  -0.112 &s    \\ 
121& 03:31:47.90& -27:48:31.0&23.85& 0.270&  5.6 & J033147.91-274831.2&   23.42 & 23.07  &-0.11  & 0.57  &  0.07 & 32231& G  &0.652 &NELG   &	& $<$5.13e-16  &  $<$41.91 &  42.49  &$<$-0.587&e,k  \\
122& 03:31:59.51& -27:50:21.7&23.89& 0.224&  3.8 & J033159.53-275021.6&   23.66 & 23.53  & 0.08  & 0.16  &  0.23 & 28406& G  &1.574 &Gal    &	& $<$5.71e-16  &  $<$42.9  &  43.25  &$<$-0.354&g,k  \\
125& 03:31:47.94& -27:50:45.5&23.91& 0.292&  6.0 & J033147.98-275045.5&   23.78 & 23.61  & 0.34  &-0.27  &  0.56 & 27530& G  &1.065 &BLAGN  &170& $<$1.56e-15  &  $<$42.92 &  42.80  &$<$0.114 &d    \\
126& 03:32:35.37& -27:49:20.6&23.92& 0.261&  4.5 & J033235.38-274920.8&   23.78 & 23.51  & 0.16  & 0.03  &  0.15 & 30561& G  &0.666 &Gal    &	& $<$6.00e-16  &  $<$42    &  42.34  &$<$-0.342&f,k  \\
130& 03:32:28.65& -27:38:46.5&23.95& 0.250&  4.0 & J033228.65-273846.8&   24.02 & 23.57  & 0.01  & 0.29  &  0.10 & 52921& G  &0.383 &NELG   &	& $<$9.66e-16  &  $<$41.63 &  41.74  &$<$-0.112&g,k  \\
131& 03:32:35.36& -28:00:41.2&24.00& 0.357&  5.7 & J033235.38-280041.4&   24.07 & 23.76  & 0.13  & 0.38  &  0.02 &  7139& Q/G&3.160 &BLAGN  &408&    2.22e-15  &     44.23 &  43.90  &    0.325&s    \\ 
\hline
\end{tabular}
$^a$ COMBO-class: G=galaxy, Q=QSO, S=star, U=unclear \citep{wolf04}.\\
$^b$ Spectroscopic class: BLAGN=Broad Line AGN, NELG=Narrow Emission Line Galaxy, Gal=galaxy.\\
$^c$ Observed in our spectroscopic follow-up.\\
$^d$ Spectroscopic redshift from \cite{szok04}, $^e$ from \cite{ravi07}, $^f$ from \cite{graz06}, $^g$ from \cite{le-f04}, $^h$ from \cite{pope08} $^s$ from \cite{trei08}. \\
$^k$ Our estimate from the CDFS intensity maps; $^l$ our estimate from the ECDFS intensity maps\\
$^m$ Extra-nuclear \citep{lehm06}.\\
$^n$ XID from \citet{giac02}.\\
$^p$ Detected in radio by \cite{kell08}.\\

\vspace{5cm}

\caption{Summary of variability and X-ray  results in the area with X-ray information.} 
\label{tab3}      
\begin{tabular}{lccccc}
\hline\hline                 
& galaxies & NELG & BLAGN & NLAGN$^a$ & total \\    
\hline                        
Variable sources detected in hard X-rays          & 1  & 3  & 42 & 0  & 46 \\
Variable sources not detected in hard X-rays      & 4  & 6  & 5  & 0  & 15 \\
\hline
Total number of variable sources                  & 5  & 9  & 47 & 0  & 61 \\
\hline
Non variable sources detected in hard X-rays      & 10 & 36 & 13 & 15 & 74 \\
Non variable sources not detected in hard X-rays  & 11 & 18 & 5  & 3  & 37 \\
\hline
Total number of non variable sources              & 21 & 54 & 18 & 18 & 111\\
\hline
\hline
Total                                             & 26 & 63 & 65 & 18 & 172\\
\hline
\hline                                   
\end{tabular}

$^a$ NLAGN=Narrow Line AGN, reported by \cite{szok04} as high excitation line sources (HEX) and by \cite{trei08} as obscured AGN (OAGN).\\
\end{minipage}

\end{sidewaystable*}